\documentclass[12pt,preprint]{aastex}

\def\cm-2{cm$^{-2}$}

\def\chandra{{\sl Chandra }}

\def\xmm{{\sl XMM-Newton }}
\def\n253{\object{NGC~253}}
\def\m33{\object{M33}}
\def\mx7{\object{M33~X$-$7}}
\def\x7{\hbox{X$-$7}}

\slugcomment{}
\shorttitle{XMM-Newton Observations of Two Candidate Supernova Remnants}

\begin{document}

\title{XMM-Newton Observations of Two Candidate Supernova Remnants\footnote{Based on observations obtained with XMM-Newton, an ESA science mission with instruments and contributions directly funded by ESA Member States and NASA.}}
\author{O. Kargaltsev$^1$, B. M. Schmitt$^2$, G. G. Pavlov$^{2,3}$, Z. Misanovic$^4$}
\affil{
$^1$Dept. of Astronomy, University of Florida, Gainesville, FL
32611-2055, USA \\ 
$^2$Dept.\ of Astronomy and Astrophysics,
The Pennsylvania State University, \\
525 Davey Lab., University Park,
PA 16802\\
$^3$St. Petersburg State Polytechnical University, Polytechnicheskaya ul. 29,
St.\ Petersburg 195251, Russia\\
$^4$School of Physics, Monash University, Melbourne, 3800 VIC,
  Australia}

\begin{abstract}
Candidate supernova remnants G23.5+0.1 and  G25.5+0.0   were observed by {\sl XMM-Newton} in the course of a snap-shot survey of plerionic and composite SNRs in the Galactic plane.
In the field of G23.5+0.1, we detected an extended source, $\sim$3$^{\prime}$ in diameter, which we tentatively interpret as a pulsar-wind nebula (PWN) of the middle-aged radio pulsar B1830--08 (J1833--0827; $P$ $=$ 85.3 ms, $\tau$ $=$ 147 kyr, $\dot{E}$ $=$ 5.8 $\times$ 10$^{35}$ erg s$^{-1}$, $d=$ 5.7 kpc), with the PWN luminosity $L_{\rm0.2-10\,keV}$ $\approx$ 5 $\times$ 10$^{33}$ erg s$^{-1}$ $\approx$ 8 $\times$ 10$^{-3}$ $\dot{E}$. The pulsar is not resolved in the EPIC images. Our analysis suggests an association between PSR B1830--08 and  the  surrounding diffuse radio emission.
  If the radio emission is due to the SNR,  
  then the pulsar must be significantly younger than its characteristic age.
   Alternatively, the radio emission may come from a relic PWN.  The field also contains SGR 1833--0832 
    and another middle-aged pulsar B1829--08 [J1832--0827; $P$ $=$ 647 ms, $\tau$ $=$ 161 kyr, $\dot{E}$ $=$ 9.3 $\times$ 10$^{33}$ erg s$^{-1}$, $d=$ 4.7 kpc], none of which are detected in our observation. 
 In the field of G25.5+0.0, which contains the extended TeV source HESS J1837--069, we detected the recently discovered young high-energy pulsar J1838--0655 ($P$ $=$ 70.5 ms, $\tau$ $=$ 23 kyr, $\dot{E}$ $=$ 5.5 $\times$ 10$^{36}$ erg s$^{-1}$) embedded in a PWN with extent of 1.3$^{\prime}$. The unabsorbed pulsar\,+\,PWN luminosity is $L_{\rm2-11\,keV}$ $\approx$ 2 $\times$ 10$^{34}$ erg s$^{-1}$ $\approx$ 4 $\times$ 10$^{-3}$ $\dot{E}$ at an assumed distance of 7 kpc. We also detected another PWN candidate (AX J1837.3--0652) with an extent of 2$^{\prime}$ and unabsorbed luminosity $L_{\rm2-10\,keV}$ $\approx$ 4 $\times$ 10$^{33}$ erg  s$^{-1}$ at $d=7$ kpc. The third X-ray source, located within the extent of the HESS J1837--069, has a peculiar extended radio counterpart, possibly a radio galaxy with a double nucleus or  a microquasar. 
 We did not find any evidence of  the SNR emission in the G25.5+0.0 field.  We provide detailed multiwavelength analysis and identifications of other field sources and discuss robustness of the G25.5+0.0 and G23.5+0.1 classifications as SNRs.  
\end{abstract}


\section{Introduction}
Modern X-ray observatories
 are powerful tools for studying diffuse emission in crowded regions of the 
Galactic plane.  The diffuse X-ray emission is often associated 
with the end products of stellar evolution such as supernova remnants (SNRs) and  young 
  pulsars (PSRs) enveloped by pulsar wind nebulae (PWNe). Although 
  growing,  the number of SNRs and pulsars 
  found in 
   X-ray  
data is still significantly smaller that the corresponding numbers of 
   detections  in the radio.  For instance, only about 100 out of   
  $\sim1900$ non-recycled, rotation-powered radio pulsars 
have been detected in X-rays. Similarly, the number of known radio SNRs stands at  274 (Green 2009), while  the number of Galactic  SNRs detected in X-rays is 
 about 50 (e.g.,  
 Seward et al.\  2010).
 Therefore, further detections of SNRs and PSRs/PWNe in X-rays would be a valuable addition to the existing limited sample. However, detecting and 
 identifying sources of diffuse X-ray emission in the Galactic plane
 is 
 often
  challenging because of the large intervening absorption column  and  
  scores of 
 background/foreground point sources.  
    For instance, it is very likely that among 20,837 extended 
    emission 
     detections in the Second  {\em XMM-Newton} Serendipitous Source Catalogue (2XMM\footnote{See \url{http://xmmssc-www.star.le.ac.uk/Catalogue/xcat\_public\_2XMM.html}}) there are quite a few  SNRs and PWNe,  yet many of  them could not be 
  identified/classified automatically based solely on X-ray data.  A thorough  multivawelength analysis  is often required to understand the nature of the 
  observed extended X-ray emission.  
    To expand the current sample of X-ray observed pulsars and PWNe and establish the characteristics of the population as a whole, we have conducted an 
{\sl XMM-Newton} survey of 14 plerionic and composite SNRs or SNR candidates within an 8 kpc distance.  First results have been presented by
 Misanovic  et al.\ (2010).
    Here we report  the results of  our \xmm\ observations  of  two SNR candidates from that sample: G23.5+0.1 and G25.5--0.0.

G23.5+0.1 is a candidate SNR discovered in a 19 ks {\sl ASCA} observation
 carried out as part of the Galactic plane survey (Yamauchi et al.\ 2002). 
A radio pulsar, B1830--08
(also known as PSR J1833--0827) is located
at $l=23.386^\circ$, $b=0.063^\circ$,
 close to the center of the diffuse  X-ray emission seen in the  {\sl ASCA} image (Ueno 2005). 
It is a middle-aged ($\tau\equiv P/2\dot{P}=148$ kyr), 85 ms pulsar at 
a distance\footnote{Here and below  the pulsar distances are inferred from 
the pulsar's dispersion measure (DM) and the model of Galactic electron density by Taylor  \& Cordes (1993).} of about 5.7 kpc, with a spin-down energy loss 
rate $\dot{E} = 5.8\times 10^{35}$ erg s$^{-1}$. The pulsar is known to 
exhibit strong glitches (Hobbs et al.\ 2004), and it has the proper motion of 
 $34\pm 6$ mas yr$^{-1}$ 
 toward the Galactic north, 
away from the Galactic plane (Hobbs et al.\ 2005).  
Located nearby ($\sim 24'$ southeast)
 are 
 the radio-bright shell SNR W41 (G23.3--0.3)  and  an extended TeV source 
HESS J1834--087 (with a radial extent of  $\sim5.4^{\prime}$) projected  within the
shell. 
Clifton \& Lyne (1986) mentioned W41 as a possible host SNR to PSR B1830--08. Gaensler \& Johnston (1995) also attempted to associate the pulsar 
with W41, but concluded that the pulsar's age does not agree with the age 
of the remnant. Aharonian et al.\ (2006) discussed a possible relation between 
HESS J1834--087, W41, and PSR B1830--08 and suggested a 
``compelling positional agreement'' between HESS J1834--087 and W41. 
However, the large separation 
 between PSR B1830--08 and HESS J1834--087 
made the association questionable (Aharonian et al.\ 2006).   
  More likely, PSR B1830--08 could be associated with G23.5+0.1 if it is indeed
an SNR. 
Another middle-aged pulsar, B1829--08 
(also known as J1832--0827; $P = 647$ ms, $\tau = 161$ kyr, $d\sim 4.7$ kpc), with a  rather low 
$\dot{E}= 9.3\times 10^{33}$ erg s$^{-1}$, is located $\sim16'$ southwest of 
G23.5+0.1. There may  also  be a compact TeV source,  HESS J1832--084 
({\em Suzaku} AO-5 program, ObsID 506021010; PI G.\ P\"{u}ehlhofer), spatially coincident but not necessarily associated  with this pulsar.  
Finally, the field also contains a recently discovered  Soft Gamma Repeater,
 SGR J1833--0832 (Gelbord et al.\ 2010).

G25.5+0.0 was detected with {\sl ASCA}
 as an extended X-ray source (AX\,J1838.0--065) 
and classified as a possible non-thermal
 SNR by Bamba et al.\ (2003),  
who estimated the distance and diameter to be 7.8 kpc and 27 pc, respectively. 
The  TeV source HESS\,J1837--069 
 was found within 
the extent of the G25.5+0.0 SNR candidate (Aharonian et al.\  2006). Subsequently,  
Gotthelf \& Halpern (2008; hereafter GH08) studied  possible association 
between HESS J1837--069 and AX J1838.0--065 using {\sl RXTE} and archival 
\chandra data. The analysis of the {\sl Chandra}  ACIS images revealed two 
 extended X-ray sources: one  coincident with the aforementioned AX J1838.0--065, and the other (fainter and more extended) coincident with 
another {\sl ASCA} source, AX J1838.3--062. 
Both sources were found to be relatively close to the HESS source position; 
however, only AX\,J1838.0--065 fell within the $1\sigma$ extent of the 
TeV emission.  Timing analysis of the {\sl RXTE } data revealed a 
70.5 ms, young ($\tau=23$ kyr) and powerful 
($\dot{E} = 5.5\times10^{36}$ erg s$^{-1}$) pulsar J1838--0655 in the core of   AX J1838.0--065, thus
solidifying the association between  AX J1838.0--065 and the HESS source.  
 Spectral analysis of 
the {\sl Chandra} ACIS data revealed an unusually hard, strongly absorbed spectrum 
(photon index $\Gamma=0.7\pm0.2$,
 $N_H=(4.5\pm0.75)\times 10^{22}$ cm$^{-2}$, for the absobed power-law (PL) 
model).  
 The  {\sl Chandra} images also showed that the pulsar is accompanied by a 
  PWN 
   of a comparable luminosity,
 with a  much softer spectrum ($\Gamma=1.6\pm0.45$). 
  GH08 noticed that although PSR J1838--0655 is an obvious counterpart to HESS J1837--069,  the source AX J1838.3--062, tentatively classified as a candidate  PWN, could also contribute to the TeV emission seen from HESS J1837--069.  Spectral analysis of the ACIS data on AX J1838.3--062 gave poorly constrained 
parameters: $0.7 < \Gamma < 3.6$ and $N_H=(7\pm5)\times 10^{22}$ cm$^{-2}$, for the absorbed PL model fit.

Here we present an analysis of \xmm observations of these two 
 SNR candidate fields rich with interesting high-energy sources. 
In Section 2 we describe the observations. Section 3 provides a summary 
of the data analysis and source detection techniques 
 and describes the properties of the detected X-ray sources.
 In Section 4 we 
 discuss the classification of the detected X-ray sources, based on optical, 
near-infrared, and high-energy data.  Finally, we discuss the pulsar/PWN 
candidates detected in our observations in Section 5
 and summarize our conclusions in Section 6. 

\section{Observations}

The fields of the SNR candidates G23.5+0.1 and G25.5+00
 were observed with the \xmm EPIC PN and MOS
detectors in Full Frame mode with Medium filter.  This mode offers the
time resolution of 73.4 ms for PN and 2.6 s for MOS.  
The G23.5+0.1 field was observed in a single pointing, while 
the G25.5+0.0 field was covered with two
 pointings (with a $9'$ offset).   Additional details of these observations are listed in Table 1.  
  There were no
strong background ßares during these observations.

\section{Source Detection and Data Analysis}

The data were reduced and analyzed using the  \xmm Science Analysis System 
(SAS), ver.\ 8.0.1.  Calibrated event files for the PN, MOS1, and MOS2 detectors  were produced using 
  the SAS tasks {\em epchain} and {\em emchain}, following standard procedures.  
To search for X-ray sources 
 and extract their properties, 
  we used the SAS task {\em edetect\_chain}. This task 
  runs on the event lists and invokes several other SAS tasks to produce 
background, sensitivity, and vignetting-corrected exposure maps.
 Three image sets (one per detector), three event lists (one per detector), 
and a user-defined likelihood threshold parameter {\em mlmin} for each of 
the fields 
  were used as inputs. In both fields we set the {\em mlmin} 
parameter to 7, corresponding to a $\simeq 4\sigma$ detection. 
 
In addition, we produced  combined 
 (mosaicked)  PN, MOS1, and MOS2 images (using the SAS task {\em emosaic}),  
smoothed with a gaussian kernel ($r = 24''$). We 
visually inspected the combined, smoothed images to  
look for
 faint extended emission, which  may not 
  be  detected automatically by {\em edetect\_chain} because of 
 algorithm limitations.

 Since the fields of view (FOV) of the two observations  
 of G25.5+0.0 
are partly overlapping,  we searched for X-ray sources in each of the two
pointings as well as  in the combined data.
 This allowed us to look for a long term variability of the sources located in the overlapping region.

Given the short exposure times, spectral analysis is warranted only for
 relatively bright 
 sources.
We chose the signal-to-noise (S/N) ratios of 25 and 10  
as thresholds  
for point and diffuse sources, respectively. There are only two sources 
(1 and 8) in the G23.5+0.1 field  and four sources (1, 6, 7, and 15) in the  
G25.5+0.0 field that meet this requirement. The SAS spectral extraction 
metatask {\em especget} invokes several other tasks to produce the source 
spectra, background spectra, response files, and effective area files. 
We maximized the S/N by extracting spectra from all the detectors 
whenever possible. 
   The spectra were binned  and then fitted with XSPEC (ver.\ 12.4.0ad). 
Two spectral models were used to fit the spectra: the optically-thin  
   thermal plasma emission model ({\em mekal}) was used for soft
sources (in which most of the emission is at energies $\lesssim 3$ keV),
 while the 
 PL 
   model
    was used for sources exhibiting harder spectra.  For each source we quote the measured, absorbed fluxes (based on the best-fit model)  for the energy ranges where there are enough counts.  We also used the XSPEC task {\em cflux} to estimate unabsorbed X-ray fluxes (see Tables 4 and 5).

Keeping in mind that most sources are too faint for spectral fitting,   
 we crudely estimated their X-ray fluxes from count rates  using the same
 energy conversion factors (ECFs) as in the 2XMM catalog\footnote{See
                                http://xmmssc-www.star.le.ac.uk/Catalogue/2XMM/UserGuide\_xmmcat.html\#EmldetFit },
  which were calculated for an absorbed PL spectrum with $\Gamma=1.7$  and $N_H=3\times10^{20}$ cm$^{-2}$.
 The low $N_H$ is appropriate for  about $30\%$ of the detected sources (which are likely to be nearby foreground stars, see below)  while  other sources  likely to have larger  $N_H$ 
 varying significantly from source to source. Therefore, for some heavily absorbed sources the fluxes provided in Tables 2 and 3 may overestimate the actual flux  by up to a factor of  3.  
  However, we use these fluxes solely for source classification purposes and  our classification scheme is rather insensitive to X-ray flux variations at such a level  (see \S5 and Fig.~7).   For several bright sources we perform the actual spectral fits  and provide accurate fluxes (see below),  which differ from those in Tables 2 and 3 by up to a factor of  3 due to a the pre-defined, fixed model being used in calculating the ECFs.  
  For   each  source we also estimated the hardness ratio,   
 \begin{equation}
   {\rm HR} =(f_{2-12}  -  f_{0.2-2})/(f_{2-12}  +  f_{0.2-2})\,,
\end{equation}
 where $f_{0.2-2}$ and  $f_{2-12}$ are the  observed (absorbed)
   fluxes  in the 0.2--2 and 2--12 keV bands, respectively.
    The hardness ratios given in Tables 2 and 3 are used for source classification together 
   with the multiwavelength properties.

\subsection{G23.5+0.1 field}

Excluding 
 spurious sources along the chip gaps and bad columns,  the  {\em edetect\_chain} task  
 detected  7 sources in the G23.5+0.1 field.
   These sources are listed in 
Table 2   and shown in Figure 1. All  the 
  automatically detected    sources  in the G23.5+0.1  field 
  appear to be point sources. 
In addition,  by
visual inspection we  
 found a region of diffuse X-ray emission  (labeled Source 8) at the
center of the G23.5+0.1 field (see Figure 1). Source 8 was not automatically detected (the
detection algorithm is not sensitive to faint  diffuse sources having large extent), 
  but the diffuse
emission, mostly confined within a $2'$ radius region,
  is clearly visible, especially in the 
smoothed image 
(Figure 1, middle).

\subsubsection{Spectral fits for brighter sources}

{\em Source 1:}  The spectrum 
  was extracted 
  from a  
 $20''$ radius aperture, for the PN, MOS1, and MOS2,
    resulting in  $\approx 1200$
  combined, background-subtracted counts. The background spectrum was extracted from a nearby 
  circular 
   region  of  80$^{\prime\prime}$ radius. The counts were grouped in 65 PN, 23 MOS1, and 19 MOS2 energy bins with about 20 total counts per bin.
The spectrum  does not fit the absorbed PL model
  but fits 
 well
 the  absorbed  {\em mekal} model with 
  $kT = 0.6\pm0.2$ keV and $N_H = (1.3\pm0.2)\times 10^{21}$ cm$^{-2}$ 
($\chi_\nu^{2} = 1.2$ for 71 degrees of freedom [d.o.f., hereafter]), 
 suggesting  X-ray emission from a foreground star 
   (see Section 4). 
    The observed (absorbed) X-ray flux in the 0.3--5 keV is 
$f_{\rm0.3-5} = (1.8 \pm 0.1)\times10^{-13}$ erg s$^{-1}$ cm$^{-2}$. 
 The detailed results of the spectral fitting are given in Table 4, and the fit is shown 
   in Figure 3.

{\em  Source 8:}  The spectrum of the extended emission at the center of 
G23.5+0.1 (Source 8 in Figure 1) was extracted from a circular aperture with the 
radius of $70^{\prime\prime}$ from the PN, MOS1, and MOS2 images. This gave 
1565 combined, background-subtracted counts (background contribution $\simeq35\%$). The background spectrum was 
extracted from a nearby circular  region  of  the $80''$ radius.
  The counts 
were grouped in 20 PN, 22 MOS1, and 22 MOS2 energy bins with about 70 total counts per bin in PN and 22 total counts in MOS1 and 2. The spectrum fits best the absorbed PL model ($\chi_\nu^{2}= 1.3$ for 48  d.o.f) 
with photon index $\Gamma= 2.3\pm0.8$ and $N_H = (3.9\pm1.9)\times 10^{22}$ 
cm$^{-2}$. The large hardness ratio,  HR=$0.9\pm0.1$, reflects the fact that the source is very strongly absorbed even though it has a rather soft  spectrum.  The observed absorbed X-ray flux in the 1--10 keV band is $f_{\rm1-10} = 
(2.0 \pm 0.3)\times10^{-13}$ erg s$^{-1}$ cm$^{-2}$. The detailed results of the spectral fitting are given in Table 5, and the fit is shown 
    in Figure 4. We have also 
 analyzed archived data from an 8 ks {\sl Chandra} ACIS observation 
 of the G23.5+0.1 field
(PI B.\ Gaensler; ObsID  10524).  
      In the ACIS-S3 image (see the inset in Fig.~1, middle panel) we found a region of faint diffuse emission and 
 a point source, CXOU~J183340.3--082830 (located $\approx58''$ north of the radio position of  PSR B1830--08), both of  which are located within 
     the extent of  the diffuse Source 8.

\subsubsection{Upper limits on undected sources}
As we did not detect X-ray emission from PSR B1830--08,  PSR B1829--08, 
and SGR J1833--0832, we can only estimate upper limits on their
fluxes. The upper limit on the flux 
 depends on the
assumed spectral shape. Spectra of middle-aged pulsars, such as
B1830--08 and B1829--08, can be usually described by a sum of PL and
blackbody components. Taking as a proxy the spectrum of the nearby
middle-aged ($\tau=110$ kyr) pulsar B0656+14 (De Luca et al.\ 2005),  with the  hydrogen column density changed from $4\times10^{20}$ to $3\times10^{22}$  cm$^{-2}$,  
we obtained   
upper limits of  $4\times 10^{-14}$  and 
$3\times10^{-14}$ ergs s$^{-1}$ cm$^{-2}$  (at 90\% confidence) 
on the observed flux in the 0.2--10 keV band  for  PSR B1830--08 and  PSR B1829--08, respectively\footnote{The limits, estimated with PIMMS; \url{http://heasarc.gsfc.nasa.gov/Tools/w3pimms.html}),  taking vignetting into account.}. 
Similar upper limits are obtained assuming an absorbed PL model with $\Gamma=2$ and the same $N_H$.  
 The upper limits on the unabsorbed fluxes are $6\times 10^{-13}$ and 
$4\times10^{-13}$  ergs s$^{-1}$ cm$^{-2}$, respectively.
 
SGR J1833--0832 was in a quiescent state during our observation.
   It is not obvious which spectral model should be assumed for a magnetar 
in a quiescent state
 because the spectra
 are different in different sources, and, apparently, some quiescent magnetars are simply 
 too dim to measure their spectra.  
If the best-fit spectral model\footnote{A two-component black-body with $kT_{\rm cold}\approx167$ eV, $R_{\rm cold}\approx9.3$ km,  $kT_{\rm hot}\approx330$ eV, and $R_{\rm hot}\approx 0.9$ km (see Table 3 in Bernardini et al. 2009).} of the transient 
 magnetar XTE J1810--197 
  in its  
lowest flux state (2007 September) is taken as a proxy,  an upper limit   
$f_{0.2-10} < 6\times10^{-15}$ ergs s$^{-1}$ cm$^{-2}$ 
(at 90\% confidence 
in 0.2--10 keV) on the 
 flux  of  SGR J1833--0832 is obtained 
(the corresponding unabsorbed flux is 
$f_{0.2-10}^{\rm unabs}\lesssim 1.6\times10^{-13}$ ergs s$^{-1}$ cm$^{-2}$ for  $N_H=3\times10^{22}$ cm$^{-2}$, close to that estimated by Esposito et al.\ 2011).   If, instead, the spectrum 
  of the quiescent
     SGR 0526--66  is taken as a proxy  
(PL with $\Gamma=3.3$; Tiengo et al.\  2009),  
we obtain similar limits:
 $f_{0.2-10}\lesssim 5\times10^{-15}$ ergs s$^{-1}$ cm$^{-2}$ 
 and $f_{0.2-10}^{\rm unabs}\lesssim1.4\times10^{-13}$ ergs s$^{-1}$ cm$^{-2}$,  
again assuming $N_H=3\times10^{22}$ cm$^{-2}$.
  We note that these estimates are approximate and may vary by a factor of a 
few depending on the actual spectrum of SGR J1833--0832 in quiescence. 
The reason 
the limits are  shallower for the pulsars  compared to  SGR J1833--0832 
is that  PSR B1830--08 is located within the enhanced diffuse emission region,
 and   PSR B1829--08 is near the edge of the EPIC 
 FOV (see 
Figure 1), where the sensitivity is a factor of 2.5 lower. Note, however, 
that the fits to the SGR J1833--0832 spectrum in its active state indicate 
an absorbing column $N_H=(1.0$--$1.6)\times10^{23}$ cm$^{-2}$ 
(G\"{o}\u{g}\"{u}\c{s} et al.\  2010),  which  
 is an order of magnitude higher then the total HI column in that direction
($1.7\times10^{22}$ cm$^{-2}$).
 If this extra absorption 
is not 
 intrinsic to the active state, then the source must 
be located behind a very dense, cold molecular cloud, 
 in which case the upper limit on the absorbed  flux would go 
up by a factor of 2--2.5 (more for the unabsorbed flux limit),
 thus bringing it close to  the $3.4\times 10^{-13}$ erg s$^{-1}$ cm$^{-2}$ upper limit reported  by G\"{o}\u{g}\"{u}\c{s} et al.\  (2010) from the 8 ks {\sl Chandra} 
ACIS observation.

\subsection{G25.5+0.0 field}

Automated  {\em edetect\_chain}  detection procedure reported 23 reliable detections (excluding spurious detections along the chip gaps and bad columns). These sources are   shown in Figure 2 and listed in Table 3.
 Among these sources seven were within the FOV of both pointings (these sources are marked with an 
asterisk  in Table 3 and throughout the text). 

Three sources (1, 6, 15) were flagged as extended by the
{\em edetect\_chain}  detection procedure, which we  confirmed by visual inspection.
Source 1, the brightest X-ray source in the field,  is  located within the $1\sigma$  extent of the TeV source  HESS J1837--069 
(see Fig. 21 in Aharonian et al.\ 2006). 
  Source 6  is either diffuse  or  multiple
   (see below).
  Extended Source 15 
  appears to be  
 elongated along
 the North-South axis on arcminute scale. 
A smoothed, mosaicked, and vignetting-corrected image (Figure 2, middle panel) was
examined for additional sources of extended emission. Upon visual inspection,
we found an extended ($\sim4'$ in diameter) 
 partial shell of faint emission near the western boundary of the FOV (indicated by a yellow ellipse in Figure 2), which could be either truly diffuse or 
 produced by multiple faint point sources clustered in in a circular shape.
 
 We note that several additional point sources, which
 were neither apparent in separate MOS/PN images  nor detected automatically,
 are seen in the combined, smoothed image. 
  Since these sources are faint, 
  little  information can be extracted 
 from the existing data even if some of 
  these sources are real. Therefore, we do not  include these sources in any of  the tables and do not
 discuss them below.
 
 \subsubsection{Spectral fits for brighter sources}
 
{\em Source 1:}
The position of the extended  Source 1 coincides with the position of 
AX J1838.0--065, detected
in {\sl Chandra} data and identified as PSR J1838--0655 
  with a PWN resolved out to $\simeq40''$--$50''$ from the pulsar (GH08). 
According to {\em edetect\_chain},  the radial extent of the source is 
$\simeq 30''$. 
 As this source happens to be outside  the FOV of the PN and MOS1 detectors, 
the spectral extraction was only possible from the MOS2 detector 
(and only in one of the two pointings), resulting in $\sim660$  background-subtracted counts 
in a $40''$ radius aperture.
 The background spectrum was extracted from a nearby circular region 
   with the radius of $80''$. The counts were grouped into 50 energy bins with about 15 counts per bin.
 We found the best-fit model to be an absorbed PL ($\chi_\nu^{2} = 0.7$
 for 41 d.o.f.) with $\Gamma = 1.25^{+0.30}_{-0.14}$ and $N_H = 5.2^{+1.0}_{-0.8}\times 10^{22}$ cm$^{-2}$.
  The fit is in agreement with the analysis of GH08, although our estimated 
absorption column seems to be slightly higher (GH08 found $N_H = 4.5^{+0.7}_{-0.8}\times 10^{22}$ cm$^{-2}$).
   For the PWN,  
   GH08 measured $\Gamma$ = 1.6$^{+0.4}_{-0.5}$ from the {\sl Chandra} ACIS data, which is almost the same as the value we found from the PSR+compact PWN (up to $r=40''$). 
    The observed X-ray flux in the 1--11 keV band is $f_{\rm1-11} = (5.4\pm 0.4)\times 10^{-12}$ erg s$^{-1}$ cm$^{-2}$.
   The detailed results of spectral fitting   are given  in Table 5 and the fit is shown in  Figure 4.

{\em  Source 6:} 
  This source 
   appears to be marginally extended (radial extent $r=11''$ as determined by {\em edetect\_chain}).
   However, {\sl Chandra} resolved this source into three point source which are  smeared 
 out in the EPIC images because of  the broader PSF of {\sl XMM-Newton}. As the sources coincide with the young ($\simeq 1$ Myr), strongly obscured, dense (core diameter $\simeq 20''$)  star-forming region W42 
   (Blum et al.\ 2000), 
    the detected emission is likely to come from magnetically active 
      and/or  pre-main-sequence (PMS) stars. 
 The spectrum of Source 6 was extracted from the PN, MOS1, and MOS2 detectors 
with $\approx$400
  combined, background-subtracted counts in an $r=33''$ circular aperture. The background spectrum was extracted from a nearby circular region with $r=80''$. Counts were grouped in  22 PN,  9 MOS1, and  9 MOS2 energy bins with 13--15 total counts per bin. 
We fitted the Source 6 spectrum with the absorbed {\em mekal} and PL models, and found that the {\em mekal} model fits better ($\chi_\nu^{2} = 1.05$
  vs.\ $\chi_\nu^{2} = 1.31$ for the PL model, for 52 d.o.f.). 
  The best-fit model implies  a high temperature, 
$kT = 2.2^{+0.4}_{-0.3}$ keV, 
 and large $N_H$ = $(2.8\pm0.4)\times 10^{22}$ 
cm$^{-2}$, consistent with the large extinction found toward W42 
($A_V\sim32$; Blum et al.\ 2000). 
  The observed X-ray flux in the 1--8 keV band  is $f_{\rm1-8}= (3.6 \pm 0.2)\times10^{-13}$ 
 erg s$^{-1}$ cm$^{-2}$.  Additional details on  spectral fitting   are given  in Tables 4 and 5, and the fits are shown in  Figures 3 and 4.

{\em  Source 7$^{*}$:}
The spectrum 
  was extracted from the PN detector only (due to a 
   low S/N ratio for the MOS data), with $\sim340$ 
 background-subtracted counts in a circular aperture with $r=20''$. 
The background spectrum was extracted from a nearby region with an
 $80''$ radius. 
  The counts were grouped into 21 energy bins with about 19 total counts per bin. The best-fit model is an 
absorbed {\em mekal} ($\chi_\nu^{2} = 0.99$ for 21 d.o.f) with 
   $kT = 0.28\pm0.07$  keV and 
$N_H = 3.8^{+1.7}_{-0.5}\times10^{21}$ cm$^{-2}$. 
The observed X-ray flux in the 0.3--2 keV band is 
$f_{\rm0.3-2} = (6.5\pm0.5)\times 10^{-14}$
  erg s$^{-1}$ cm$^{-2}$. 
The details of the spectral fit are provided in Table 4 and the spectrum  is shown in Figure 4.

{\em Source 15:} 
 This source coincides with the PWN candidate  AX J1838.3--062 (GH08). 
It appears to be extended, with the radial extent of $\approx 20''$ 
(according to {\em edetect\_chain}). A very faint arcminute-scale emission 
is seen upon visual inspection in the smoothed combined PN image. 
      The spectrum,  
  extracted from the PN, MOS1, and MOS2 detectors, 
 has $\simeq250$
  combined, background-subtracted counts in a circular
aperture with $r=60''$. The background spectrum was extracted
from a nearby circular  region  with $r=
80''$. Counts were grouped into 20 PN, 11 MOS1, and
 11 MOS2 energy bins,  with 13-14 and 6--8 total counts per bin for PN and MOS1/2, respectively.  The spectrum fits
 best  
an absorbed PL model
($\chi_\nu^{2}= 0.95$ for 45 d.o.f) with $\Gamma=2.1^{+0.9}_{-0.6}$ and $N_H =
6.7^{+3.8}_{-2.2}\times 10^{22}$ cm$^{-2}$. Thus, our fit is better constrained
than that reported by GH08 (see also Section 1). The observed X-ray flux in the 2--10 keV band is
$f_{2-10} = (3.7\pm0.9)\times 10^{-13}$ erg s$^{-1}$
cm$^{-2}$.
  As noted by GH08, the spectrum of this source
 is probably dominated by the nebular emission. The complete results  of the spectral fitting are given in Table 5, 
 while the spectrum and the fit are shown in Figure 4. 

\section{Multiwavelength Analysis}
 We use the positions of the 
  reliably detected X-ray sources (Tables 2 and 3) to perform multiwavelength cross-correlation and analysis. 
We carried out 
the cross-correlation search with the following catalogs:
 MAGPIS 
 (White et al.\  2005),
  NVSS 
 (Condon et al.\ 1998), and ATNF Pulsar Catalog (Manchester et al.\ 2005) in
the radio;
 2MASS
 (Skrutskie et al. 2006), USNO-B1.0
 (Monet et al. 2003)],
and {\sl HST} Guide Star Catalog 2.3.2 in the NIR/optical;
 and high-energy 
 ({\sl CGRO} EGRET, {\sl Fermi} LAT, {\sl INTEGRAL}) catalogs,
 using the HEASARC Browse\footnote{See \url{http://heasarc.gsfc.nasa.gov/cgi-bin/W3Browse/w3browse.pl}}  and DS9\footnote{See \url{http://hea-www.harvard.edu/RD/ds9/}} Catalog Search tools.
We also examined a list of HESS TeV sources, compiled from several online 
resources\footnote{E.g., \url{http://tevcat.uchicago.edu/}; \url{http://www.mpi-hd.mpg.de/hfm/HESS/pages/home/sources/}.} and publications.

  To 
   identify optical/NIR counterparts,
      we use the X-ray source positional uncertainty,
  calculated 
 as $\sigma_{\rm pos}=(\sigma_{\rm c}^2 + \sigma^2_{\rm sys})^{1/2}$, where
$\sigma_{\rm c}$ is the centroiding (statistical) uncertainty for each individual
X-ray source, and $\sigma_{\rm sys}$ is the systematic pointing uncertainty of {\sl XMM-Newton}\footnote{The same method is used  in the \xmm Serendipitous Source Catalog, Second
Version (2XMM), see
  \url{http://xmmssc-www.star.le.ac.uk/Catalogue/xcat\_public\_2XMM.html}.}. In 
G23.5+0.1 and G25.5+0.0 observations, the $\sigma_{\rm c}$  is between 
0\farcs26 and 2\farcs3 for all the reliably detected sources. 
The systematic uncertainty $\sigma_{\rm sys}$   is taken to be 
$1''$, similar to the 2XMM catalog\footnote{See \url{http://heasarc.gsfc.nasa.gov/W3Browse/xmm-newton/xmmssc.html}.}. We  
 find $\sigma_{\rm pos}$ to be between 
1\farcs0 and 2\farcs0 for the sources in G23.5+0.1, and
between 1\farcs1 and 2\farcs5 for the G25.5+0.0 sources.
  We consider optical/NIR sources within $3\sigma_{\rm pos}$ from the X-ray source as possible counterparts.
  The probability of  the association  
   depends on the offset $r$ between the X-ray source and its 
  NIR/optical counterpart.
 It can be estimated as 
the probability  
 of finding zero field sources in the circle of radius $r$,
$P=\exp(-\rho \pi r^2)$, 
 where  
  $\rho$  is the average surface density of the optical/NIR sources in the field.  
We measured $\rho_{\rm USNO-B1}=0.007$, 
$\rho_{\rm 2MASS}=0.015$ arcsec$^{-2}$  in the G23.5+0.1 field, 
and $\rho_{\rm USNO-B1}=0.006$, $\rho_{\rm 2MASS}=0.011$ arcsec$^{-2}$
 in the G25.5+0.0 field. The offsets and the corresponding probabilities are given in Tables 6 and 7.

\subsection{G23.5+0.1}

 \subsubsection{High Energy Counterparts} 
 
 HESS J1834--087, is located in the vicinity of G23.5+0.1 but just outside the FOV of our {\sl XMM-Newton} observations. 
 However, most of the extended TeV emission from HESS J1834--087 appears to be confined within 
  the radio
  shell  of SNR W41 (see Figure 5).
   Furthermore,
 a candidate PWN, possibly  associated with W41 and HESS J1834--087, was 
 recently reported by  Misanovic et al.\ (2011). On the other hand, the asymmetry of the HESS source (see Figure 20 in Aharonian et al.\ 2006)  suggests 
 that there may be a fainter TeV source associated with  G23.5+0.1. More sensitive TeV observations are needed to confirm this hypothesis. 
 In addition,   {\sl Suzaku}  has observed  a compact TeV source HESS J1832--084, which is apparently coincident with PSR B1829--08 in the G23.5+0.1 field,  on April 8, 2011 for 40.3 ks (PI G.\ P\"{u}hlhofer).
  However, no other information on this new VHE source have been published so far.
  
  We have also searched  the {\sl Fermi} LAT  1FGL catalog (Abdo et al.\ 2010) but found no GeV source within the field of view of our {XMM-Newton} observation.  The nearby  1FGL~J1834.3--0842c is located just outside the EPIC FOV and coincides with  W41 rather than with G23.5+0.1 (see Figure 5).

 \subsubsection{Optical and Near-Infrared Counterparts}

 We found optical/NIR counterparts to all point-like X-ray sources but 
   Sources 3 and 6. We did not find a counterpart to the extended  
   Source 8 (see Table 6). We, however, found that the point source CXOU~J183340.3--082830 (see \S3.1.1), located within Source 8,  has a NIR counterpart 2MASS~18334038--0828304 ($J=15.7$, $H=14.0$, $K=13.5$).  More detailed information including magnitudes is provided in Table 6.
 
 \subsubsection{Infrared and Radio Counterparts}

Only
 Source 6 and  the diffuse Source 8
 have no IR counterparts in the   {\sl Spitzer}'s GLIMPSE 
source catalog\footnote{Available at \url{http://irsa.ipac.caltech.edu/data/SPITZER/GLIMPSE/} }  
  (see Table 6 for details).   We did not find any radio sources coincident with the X-ray point sources. 
  However, the MAGPIS image reveals  complex diffuse emission some of which is in the vicinity of B1830--08 and may be related to its PWN or the host SNR.  A  search in the ATNF pulsar catalog (Manchester et al.\ 2005) reveals PSR B1830--08, located within the extended \xmm Source 8 (in the G23.5+0.1 field) and offset by $\sim$1$^{\prime}$ from its center, and PSR   B1829--08 (near the western boundary of the EPIC FOV; see Fig.~1), which does not have an X-ray counterpart. 
  We also notice in the MAGPIS 20 cm image the region of Source 8 appears to be inside a cavity  in a large-scale diffuse radio emission (Fig.~1, middle panel) while  PSR   B1829--08 appears 
 to be surrounded by diffuse radio emission  which could be due to a PWN.

 \subsection{G25.5+0.0}

 \subsubsection{High Energy Counterparts} 

 Figure 1 shows that  Source 1 is located 
  within the 1$\sigma$ extent of 
 HESS J1837--069.
  From the analysis of the {\sl Chandra} ACIS data GH08
have concluded that AX~J1838.0--065  
  (Source 1) was the likely source of the TeV emission, but 
  suggested that AX~J1838.3--062 (Source 15) 
   can also  contribute  to 
 the  TeV  emission.  The only other X-ray source that might contribute to the TeV emission is Source 9 located within the extent of  HESS J1837--069.
 This source appears to be point-like in X-rays (both in the {\sl Chandra} and {\sl XMM-Newton} images), but it has an interesting extended radio counterpart (see \S 4.3).

The only other GeV source that falls within the  FOV of our observations is  1FGL~J1837.5--0659c located in the vicinity of  Sources 1, 9, and 15 in the G25.5+0.0 field.  Within its positional uncertainty (shown in Figure 2  at 95\% confidence), the 1FGL source does not coincide with any of these three X-ray sources (see Figure 2, middle panel). 
  However, the positional uncertainty of this source could be  underestimated since this source is marked as confused in the 1FGL catalog\footnote{Sources labeled  ``c'' (confused) at the end of the 1FGL names are found in  regions with bright and/or possibly incorrectly modeled diffuse emission (see \url{http://fermi.gsfc.nasa.gov/ssc/data/access/lat/1yr\_catalog/}).}.  Thus, 1FGL J1837.5--0659c  still could be a counterpart to Sources 1, 9  or 15, or  it may also simply be a spurious source. As the LAT statistics improves, a better positional accuracy and a search for pulsed signal  can allow one to reveal the nature of the GeV source and identify its  X-ray counterpart.
 
AX J1838.0--065 (PSR J1838--0655; Source 1 in the G25.5+0.0 field)  is the only {\sl INTEGRAL} source 
    within the FOV of  our \xmm\ observations,  with the soft-band flux of  (1.4$\pm$0.1)$\times$$10^{-10}$ erg s$^{-1}$ cm$^{-2}$ (in 20--40 keV) and the hard-band flux of (2.5$\pm$0.2)$\times$$10^{-10}$ erg s$^{-1}$ cm$^{-2}$ (in 40--100 keV), according to Bird et al.\ (2010).  

\subsubsection{Optical and Near-Infrared Counterparts}

 We found that about $50\%$ of X-ray sources lack optical and NIR counterparts  (sources 1, 3, 9, 13,
15, 16$^{*}$, 17, 18, 19, 20, and 22 in Table 7).  Only two sources lacking optical counterparts appear to have NIR counterparts (sources 11 and 21). More detailed information including magnitudes is provided in Table 7.
 
 \subsubsection{Infrared and Radio Counterparts}

 Most of the sources in this field have counterparts in the GLIMPSE catalog, except
    for sources 
 1, 9, 15 and 19 (see Table 7 for details).  We also found radio counterparts to several X-ray sources. 
 
     In the  MAGPIS catalog we found a radio counterpart 
  (G25.31987--0.09825; Helfand et al.\ 2006) 
    to  Source 4 (see Fig.~2),   
    with the offset of just $1''$ from the X-ray position
  which is much  smaller than than the restoring beam  size\footnote{We quote only the beam size here since Helfand et al.\ (2006) do note provide positional uncertainties for the sources; however,  
 we would expect them to be similar  to those in  quoted in White et al.\  (2005), i.e.,  typically about $1''$. } of $6\farcs2\times5\farcs4$. 
     The radio source has the peak 
20 cm
     flux density of $41.10\pm0.34$ mJy.

   Source 8$^{*}$ in the G25.5+0.0 field  (see Fig.~2) was also found to have a possible radio counterpart, G25.340--0.048, with  a flux of 6.7 mJy at $\lambda=6$ cm (White et al.\  2005).  
However, this radio source is offset from the X-ray source by $3\farcs5$, while the typical $1\sigma$ uncertainty of the radio position is about $1''$ (White et al.\  2005).

  Third source (in the G25.5+0.0 field) that has a radio counterpart is the star-forming region W42 (Source 6), with  
   integrated flux densities of 1.75 Jy  at 20 cm and 1.46 Jy at 6 cm. 
     Source 6 also has multiple counterparts in the Catalog of Star-Forming Regions in the Galaxy (Avedisova 2002). The color-coded color image of G25.5+0.0 (see Figure 2, bottom) reveals a good spatial correlation between Source 6 and
    the regions of  diffuse radio and infrared emission.
    Finally, the extended X-ray emission near the western boundary of the G25.5+0.0 image 
(see Section 3.2)  
  has a  diffuse radio counterpart which appears to be 
  inside the partial-shell-shaped X-ray emission. We name this diffuse source G25.25+0.28.
   We also note that Source 21, although lacks a compact radio counterpart, is embedded into a region of faint diffuse emission apparently associated with another star-forming region, which  is also the HII region 
  G025.4+00.0 (Paladini et al.\ 2003), harboring two  young stellar objects (YSOs; Urquhart et al.\  2009; the nearest of the YSOs is still $20''$ off the Source 21 position). 
   
 There are also two 
  bright radio sources in the G25.5+0.0 field
 (A and B; indicated by arrows in Figure 2, bottom panel) located in the vicinity 
 of HESS~J1837--069.  
   As   Source 9 is found to be approximately in the middle of  the bright  linearly extended radio Source B (integrated flux is 7 Jy at 20 cm; Helfand et al.\  2006), they  are very  likely   associated (see Fig.~2, bottom panel).  Radio source B also has very interesting morphology, 
 with a double nucleus in the middle and two extended arms (jets?) on both sides of the double nucleus (see Fig.~6). The southern arm/jet appears to be better collimated,  and it terminates at a  bright spot.  The ratio of the 6 to 20 cm fluxes suggests a non-thermal spectrum ($\alpha=0.5-1$, $F_{\nu}\propto \nu^{- \alpha}$) both for extended emission and for each of the nuclei.   This radio 
 source does  not  have any counterparts in {\sl Spitzer} 
   IRAC
   or 2MASS 
  images.  We have also examined the {high-resolution \sl Chandra} ACIS image 
 and found an X-ray counterpart (CXOU~J183751.62--064355.4; 273 counts in 20 ks  in 0.5--8 keV) coincident with the southern 
  nucleus.   The ACIS spectrum  fits well an absorbed PL model  with $N_H=(1.2\pm0.3)\times10^{23}$ cm$^{-2}$,  $\Gamma\simeq1.3^{+0.4}_{-0.3}$, and 0.5--8 keV unabsorbed flux of
  $1.1\times10^{-12}$ erg s$^{-1}$ cm$^{-2}$ (observed flux is $4.3\times10^{-13}$ erg s$^{-1}$ cm$^{-2}$ in 2--8 keV).  The X-ray flux is crudely 
  consistent with the extrapolation of the corresponding radio nucleus counterpart PL spectrum ($\alpha\simeq 0.7$). The reports of HII 
  regions in the vicinity of the radio nucleus (e.g., Avedisova 2002;  Giveon et al.\ 2005)  likely stem from the assumed association between the radio source B and  a nearby bright IR source detected by the {\sl MSX} satellite. However, higher resolution {\sl Spitzer} IRAC images show that the IR source is $11''$ northwest of the radio nucleus and hence it is not associated with the radio source B.  The strong absorphion  and the lack of an optical counterpart 
  suggest that the source is either extragalactic (hence its light has to penetrate the entire Galactic disk) or it is galactic but intrinsically absorbed (e.g., an HMXB).

\section{Source Classification}

Based on the measured X-ray properties and    multiwavelength properties extracted from catalogs, we 
  attempted to  classify  the detected sources. 
  In some cases the 
   classifications  listed in Tables 6 and 7 are only tentative, based on  the limited information available.

  The hardness ratio vs.\  flux (or count rate)  plots (analogous  to the ``color-magnitude diagram''  in the optical) are often used 
   for  classification of X-ray sources (e.g.\ Ebisawa et al.\  2005; Jonker et al.\ 2011). 
  In such diagrams  the coronally active non-degenrate stars typically occupy  the low-HR, low-flux corner and hence are easy to spot.  Figure 7 shows such a diagram for our sample. 
  
  Another commonly used classification criterion  is the  X-ray-to-optical flux ratio.
    Following  Maccacaro et al. (1988),  we
     compute 
      $\log(f_{X} / f_{V}) $ = $\log f_{X} + (m_{V}/2.5) + 5.37$, where $m_{V}$ is the apparent magnitude in the \textit{V} band.
  Galaxies, stars, AGNs and X-ray binaries, each fall into a particular range of values of $\log(f_{X} / f_{V})$ (see our Fig.~8 and Fig.~1 in  Maccacaro et al.\ 1988).
We estimate the X-ray-to-optical flux ratios for our sources (see Tables 6 and 7 and text below), and use them for source classification, together with other available information.

 The X-ray hardness ratio, which is also used for source classification, does not describe uniquely the X-ray spectrum of a source even in a simple case when the spectrum fits an absorbed PL model (e.g., one often cannot discriminate between a soft, strongly absorbed source and a hard,  
 weakly absorbed source). However, since the statistics precludes  spectral fitting for 
 faint sources, the hardness ratio still can be used to differentiate between 
  the  most common types of sources: soft, weakly absorbed 
  foreground stars
   and hard, the strongly absorbed 
   AGNs, seen through the Galactic disk, or cataclysmic variables (CVs) and X-ray binaries (XRBs) in the Galactic bulge.  Adding in  the flux and IR/NIR/optical fluxes helps to brake the degeneracy because most of strongly absorbed (apparently ``hard'') sources with low X-ray fluxes and optical/NIR/IR identifications would be CVs and RS CVns (see, e.g.,  Figure 3 of Jonker et al.\ 2011) or embedded  coronally active stars (see, e.g., Getman et al.\ 2011 and references therein).

 We also 
    used colors of  the optical, NIR and IR counterparts for source classification. 
   Since the optical colors alone are not very informative  because of the unknown extinction varying among the sources, we prefer to use  NIR and NIR-IR colors which could be used to distinguish between 
   the pre-main and main sequence stars and/or AGNs.  

Based 
 on the optical, NIR, and IR data, as well as the flux ratios and X-ray hardness ratios 
  (see Figs.\ 8  and 9), we find that the majority of the X-ray sources detected in the two SNR fields are foreground main-sequence stars or  pre-main sequence stars associated with star-forming regions or AGNs. Several objects are likely to be  background CVs or quiescent XRBs, although  
   we cannot completely rule out the possibility that some of the faint X-ray  sources with no optical or infrared counterparts may be isolated  neutron stars and pulsars.  The current population models for different classes of X-ray sources are fairly uncertain (see e.g., Ebisawa et al.\  2005; Motch et al.\ 2010 and references therein) and provide only spatially averaged estimates, while the actual distribution of different types of objects can be quite inhomogeneous throughout  the Galactic plane even on a degree angular scales.  Nevertheless we used the estimates provided in Table 7 of Motch et al.\ (2010) for the {\sl XMM-Newton} Galactic Plane Survey (XGPS)  covering 3 square degree region of the plane between $l=19^{\circ}$ and $22^{\circ}$ (i.e. nearly adjacent to our fields). Scaling down the XGPS estimates by area, we would expect to detect 5-9 AGNs, ~4 coronally active stars, and about 2 accreting binaries down to the flux level of $~5\times 10^{-14}$ erg s$^{-1}$ cm$^{-2}$. The total number of sources we detected above this flux level is 70\% larger suggesting some unaccounted population or  reflecting the uncertainties of the population estimates. One possible reason is the starforming activity associated 
 the G25.5+0.0 filed (see below) likely resulting in a larger number of young magnetically active stars
  which we detect in X-rays. Part of the  excess among the hard sources is due to the two young pulsars but there still remain 2-3 unaccounted sources  which could be NSs (see Tables 6 and 7).

   Among the detected and classified sources particularly interesting are the following. 
   \begin{itemize}
\item  Source 1 (PSR J1838--0655), in the G25.5+0.0 field, is a young pulsar   whose pulsations have been found with {\sl RXTE} (GH08).
\item   We confirm the previously suggested (GH08) interpretation  of AX J1837.3--0652 (Source 15) as a PWN based on the extended X-ray emission and a lack of a NIR/IR counterpart.
\item    Source 9 (having an extended radio counterpart) in the same field could  be a peculiar   
     radio galaxy seen through the Galactic disk or an HMXB in the low state.
 \item   A partial shell diffuse source G25.25+0.28, resolved in X-rays and radio,   is  likely to be an uncataloged SNR. 
 \item    In the  same field  we also detected  
 several  X-ray sources apparently  associated with 
 a massive young cluster  W42 (based on the location and correlations with IR sources). 
 \item  In  the G23.5+0.1  field we discovered   an 
 extended Source 8 
    which is likely an X-ray  PWN powered by PSR B1830--08. 
\end{itemize}

\section{
 Discussion and Summary}

\subsection{The field of G23.5+0.1}

In the G23.5+0.1 field we found an extended ($\sim$4$^{\prime}$ in diameter) X-ray emission surrounding PSR B1830--08, which is likely a
  PWN. Its   X-ray efficiency in 0.5--8 keV, $\eta_{0.5-8}\equiv L_{0.5-8}/\dot{E}=5\times10^{-3}$,   
 is well  within the range   
measured for other PWNe 
  (Kargaltsev \& Pavlov 2008).
  The undetected pulsar must be at least a factor of 10 less luminous than the PWN, which is not unusual.
 The extended X-ray emission has a rather amorphous shape which does not appear as a bow-shock PWN (see also Esposito et al.\ 2011). This suggests that despite the apparently high but rather uncertain transverse velocity of $751\pm163$ km s$^{-1}$     (calculated for the dispersion measure distance of 4.7 kpc; see Hobbs et al.\ 2005), PSR B1830--08 either does not move 
   supersonically (because the distance is strongly overestimated)  or  the detected extended X-ray  emission does not come from  the PWN (in which case it might be associated with an SNR). 
  The faint diffuse radio emission
    north of  PSR B1830--08 (see Fig.~1, bottom and Figure 5) could be related to the host SNR.
     The thermal X-ray emission from the SNR shell could be strongly absorbed due to the large $N_H$. If this radio emission is indeed due to the host SNR of PSR B1830--08, then the pulsar must be much younger than its spin-down age    of 147 kyrs.  This is possible, and indeed a similar age discrepancy was found, e.g.,  for PSR J0538+2817 in SNR S147 (Ng et al.\ 2007).
   Alternatively, it is possible that the diffuse radio emission comes from the relic PWN of  PSR B1830--08,  where the electrons are too ``cold'' to emit X-rays via synchrotron mechanism. A number of relic PWNe has been recently found (see Kargaltsev \& Pavlov 2010 for a review), mainly in the TeV observations. It is possible that a deeper TeV exposure would detect a relic PWN of  PSR B1830--08, a hint of which might be already seen (\S4.1). If the radio emission north of PSR B1830--08 is due to its relic PWN, 
     the host radio SNR 
 shell may have either become too dim or  too large to be seen in the X-ray and radio images. 
    Additional confusion is caused by the highly nonunifom, complex radio background due to the adjacent  W41
   southeast of  G23.5+0.1 and the radio-bright (apparently star-forming) region northeast of the pulsar (which correlates with IR emission and has a thermal spectrum  in the radio; Shaver \& Goss 1970). 
  
  We also detected seven other point sources in the 
  field of G23.5+0.1. All but one (Source 6) were found to have IR counterparts and hence 
    are likely foreground stars, CVs, XRBs, or AGNs seen through the Galactic disk (see Table 6). Source 6 has too few X-ray counts to make a conclusion about its nature, it could be an AGN or  a faint pulsar.  
 Finally, there are two possible high-energy sources which we do not detect in our observation. One is  PSR B1829--08, which has a spin-down age similar to that of PSR B1830--08 and 
   located at nearly the same distance (according to  the Galactic  electron density model by Taylor  \& Cordes 1993) 
    but  has the spin-down power 
    a factor of 62 lower  then B1830--08. 
    However, 
 the surrounding diffuse radio emission and  the good spatial coincidence with the TeV source HESS J1832--084  may indicate the presence of a relic PWN whose electrons are too cold to emit X-rays via the synchrotron mechanism.   
    We also have not detected  SGR 1833--0832  in our {\sl XMM-Newton} observation.   Assuming a plausible spectral model for the quiescent  SGR spectrum and $N_H=3\times10^{22}$ cm$^{-2}$, we find that its quiescent absorbed flux is at least a factor of $10^3$
 lower than  its active state flux. This places a very low limit on the quiescent  SGR luminosity, $L\lesssim 10^{32}(d/8~{\rm kpc})^{2}$ erg s$^{-1}$ (in 1--10 keV, unabsorbed, assuming $N_H=3\times10^{22}$ cm$^{-2}$), a factor of  4,000  smaller than that of   
 SGR 0526--66 measured 30 years after the giant flare (Tiengo et al.\ 2009). This suggests that SGRs can maintain an elevated luminosity decades after the period of activity,
 well above the ``truly quiescent'' level. One of the implications is that there may be a significant number of  quiescent SGRs, which we have not detected in X-rays so far.

\subsection{The field of G25.0+0.0}

 The only  large-scale diffuse emission we found in the  G25.0+0.0  field is a shell-like source  G25.25+0.28. However, its low surface brightness in X-rays precludes any spectral analysis.  G25.25+0.28 is too far from the center of  G25.0+0.0 to contribute to the G25.0+0.0 emission seen in the {\sl ASCA} images. 
  We conclude that the previous claims of extended emission and  
 the reports of the 
  X-ray SNR 
 candidate
   G25.5+0.0 (e.g., Bamba et al.\ 2003) were likely the result of    overlapping     {\sl  ASCA}  PSF wings from several adjacent  unresolved sources.
    One of these, Source 6, has  a thermal-like X-ray spectrum and positionally  coincides  with the core of the young,  obscured  star-forming region W42. 
  Although Source 6 appears to be point-like at the {\sl XMM-Newton} resolution, it is resolved into three sources in the {\sl Chandra} ACIS image.  
     
     {\sl XMM-Newton} and {\sl Chandra} also resolved two 
    {\em ASCA} sources, one of which (AX J1838.0--0655) is a recently discovered  young pulsar (J1838--0655) while the other one (AX J1837.3--0652) is likely a PWN powered by a yet undetected pulsar (GH08). 
    Although each of these two objects 
      could be accompanied by an SNR,  
      the radio shell could have dissipated significantly 
       (the pulsar's spin-down age is
         23 kyrs)  while the thermal X-ray emission must be severely attenuated by  the large absorbing column. 
        
          We also identified a peculiar radio counterpart to the Source 9 which could be a heavily obscured radio galaxy with powerful jets 
             (cf.\ M87 or Centuarus A)
           and double nucleus (but not a blazar since the jets are  at significant angle with respect to the observer). Similar sources (e.g., 4C 65.15 and 3C433) have been described by Miller \& Brandt (2009) and classified  as FR~II radio galaxies with asymmetric environments. 
             Alternatively,  Source 9 could be a heavily absorbed microquasar (e.g., similar to 1E 1740.7--2942) but in the very low state. Note that a similar double-lobed radio and X-ray source was recently found within TeV~J2032+4130  (Butt et al.\ 2008).  
       All the three sources could contribute to the TeV emission from HESS~J1837--069  and the possible GeV emission from the 
      {\sl Fermi}
       source 1FGL J1837.5--0659c. 
      
      The only other sources that do not have IR, NIR, or optical counterparts are Sources 13 and 19. Particularly interesting is Source 19 that   
       has HR$\simeq-0.04$, suggesting a  soft and relatively weakly absorbed X-ray spectrum that could belong to a foreground star, but the lack 
       of  any optical/NIR counterpart disfavors such a scenario.  AGNs, quiescent XRBs, and CVs are expected to have much harder spectra. A plausible remaining option is a nearby, ``X-ray dim'' isolated neutron star  (see, e.g., Kaplan 2008 for a review).  However, the scarce data available make difficult any firm conclusions. 
       Deeper optical/NIR observations are needed to make a firm conclusion on the nature of this source.
  
  \acknowledgments
{We would like to thank Richard White for the help with MAGPIS images,  Aya Bamba for providing references on ASCA observations, and Neil Brandt for the discussion of Source 9 in the G25.0+0.0 field.  We also thank Konstantin Getman for 
useful discussion about the properties of young stars.  We are grateful to the anonymous referee for very careful reading and useful suggestions.   The work on this project was partly 
supported through the NASA grants  NNX06AG36G,  NNX09AC81G and 
NNX09AC84G,  and National Science Foundation grants No.\  0908733 and 0908611.  The work by G.~G.~P.\  was partly supported by the
Ministry of Education and Science of the Russian Federation (contract 11.634.31.0001).}

\clearpage

\begin{center}
 REFERENCES:
\end{center}


\noindent Abdo, A. A., Ackermann, M., Ajello, M., et al.  2009,  ApJS, 183, 46

\noindent Abdo, A. A.,  Ackermann, M., Ajello, M., et al.  2010, ApJS, 188,  405

\noindent Aharonian, F., Akhperjanian, A. G., Bazer-Bachi, A. R., et al. 2006, ApJ, 636, 777

\noindent  Avedisova, V. S. 2002, Astronomy Reports, 46, 193

\noindent  Bamba, A., Ueno, M., Koyama, K., \& Yamauchi, S.\ 2003, \apj, 589, 253

\noindent  Bernardini, F., Israel, G. L.,  Dall'Osso, S.,   et al.  2009, \aap, 498, 195 

\noindent  Bird, A. J.,  Bazzano, A., Bassani, L., et al.\ 2010, ApJS, 186, 1

\noindent Blum, R. D., Conti, P. S., \& Damineli, A.\ 2000, \aj, 119, 1860 

\noindent Butt, Y. M., Combi, J.~A., Drake, J., Finley, J.~P., Konopelko, A., Lister, M., Rodriguez, J., \& Shepherd, D.\ 2008, MNRAS, 385, 1764

\noindent Clifton, T. R., \& Lyne, A. G.\ 1986, \nat, 320, 43 

\noindent  Condon, J. J., Cotton, W. D., Greisen, E. W., Yin, Q. F., Perley,
R. A., Taylor, G. B., \& Broderick, J. J.\ 1998, \aj, 115, 1693

\noindent  Ebisawa, K., Tsujimoto,  M., Paizis, A., et al.\ 2005, \apj, 635, 214

\noindent  Esposito, P., Israel, G.\ L., Turolla, R.,  et al.   2011, \mnras, 416, 205

\noindent Gaensler, B. M., \& Johnston, S.\ 1995, \mnras, 275, L73 

\noindent Giveon, U., Becker, R. H., Helfand, D.~J., \& White, R.L.\ 2005, \aj, 130, 156

\noindent  G{\"o}{\u g}{\"u}{\c s}, E., Cusumano, G., Levan, A. J.,  et al.\ 2010, \apj, 718, 331

\noindent Gelbord, J. M.,  Barthelmy, S. D., Baumgartner, W. H., et al. 2010, GRB Coordinates Network, Circular Service, 10526, 1 (2010), 526, 1

\noindent  Getman, K. V., Broos, P. S., Feigelson, E. D., et al.\  2011, ApJS, 194, 3

\noindent  Gotthelf, E.~V., \& Halpern, J. P.\ 2008, \apj, 681, 515  (GH08)

\noindent   Green, D. A.\ 2009, Bulletin of the Astronomical Society of India, 37, 45 (available on the World-Wide-Web at "http://www.mrao.cam.ac.uk/surveys/snrs/")

\noindent  Helfand, D. J., Becker, R. H., White, R. L., Fallon, A., \& Tuttle, S.\ 2006, \aj, 131, 2525

 \noindent Hobbs, G., Lyne, A. G., Kramer, M., Martin, C. E., \& Jordan, C.\ 2004, \mnras, 353, 1311

 \noindent Hobbs, G., Lorimer, D. R., Lyne, A. G., \& Kramer, M.\ 2005, \mnras, 360, 974 

 \noindent Jonker, P. G., Bassa, C. G., Nelemans, G., et al.\  2011, ApJS, 194, 18

\noindent  Kaplan, D. L.\ 2008, 40 Years of Pulsars: Millisecond Pulsars, Magnetars and More, AIP Conf.\ Proc.,  983, 331 

\noindent  Kargaltsev, O., \& Pavlov, G. G. 2008, in 40 Years of Pulsars: Millisecond Pulsars, Magnetars, and More, eds. C. Bassa, A. Cumming, V. M. Kaspi, \& Z. Wang, AIP Conf. Proc., 983, 171 

\noindent  Kargaltsev, O., \& Pavlov, G.\  2010,  in X-ray Astronomy 2009: Present Status, Multi-Wavelength Approach  And Future Perspectives, eds.\  A.\ Comastri, M.\ Cappi, \& L.\ Angelini,  AIP Conf. Proc.,   1248, 25

\noindent Kaspi, V.\ M. 2007, Ap\&SS, 308, 1 

\noindent De Luca, A., Caraveo, P. A., Mereghetti, S., Negroni, M., \&
Bignami, G. F.\ 2005, \apj, 623, 1051

 \noindent Maccacaro, T., Gioia, I. M., Wolter, A., Zamorani, G., \&
 Stocke, J. T. 1988, ApJ, 326, 680

 \noindent Manchester, R. N., Hobbs, G. B., Teoh, A., \& Hobbs, M.\ 2005, \aj, 129, 1993 

\noindent Miller, B. P., \& Brandt, W.~N.\ 2009, \apj, 695, 755

 \noindent  Misanovic, Z., Kargaltsev, O., \& Pavlov, G. G.  2010, ApJ, 725, 931
 
  \noindent Misanovic, Z., Kargaltsev, O., \& Pavlov, G. G.  2011, \apj, 735, 33

 \noindent  Monet, D.~G., Levine, S. E., Canzian, B., et al.\ 2003, \aj, 125, 984 

 \noindent  Motch, C., Warwick, R., Cropper, M. S., et al.\ 2010, \aap, 523, A92 

\noindent   Ng, C.-Y., Romani, R.~W., Brisken, W.~F., Chatterjee, S., \& Kramer, M.\ 2007, \apj, 654, 487

 \noindent   Paladini, R., Burigana, C., Davies, R.~D., Maino, D., Bersanelli, M., Cappellini, B., Platania, P., \& Smoot, G.\ 2003, \aap, 397, 213
 
  \noindent  Pavlov, G.\ G., Kargaltsev, O., Wong, J.\ P., \& Garmire, G.\ P.  2009, ApJ, 691, 458 

\noindent  Seward, F., Slane, P., Randall, S., et al.\ 2010,  Chandra
Supernova Remnant Catalog, Harvard-Smitonian Center for Astrophysics,
Harvard University, USA (available on the World-Wide-Web at "http://hea-www.harvard.edu/ChandraSNR/").

\noindent  Shaver, P. A., \& Goss, W. M.\ 1970, Australian Journal of Physics Astrophysical Supplement, 14, 133

\noindent  Skrutskie, M. F., Cutri, R. M., Stiening, R., et al.\ 2006, \aj, 131, 1163

\noindent  Taylor, J. H., \& Cordes, J. M. 1993, ApJ, 411, 674

\noindent Tiengo, A., Esposito, P., Mereghetti, S.,  et al.\ 2009, \mnras, 399, L74 

\noindent White, R. L., Becker, R. H., \& Helfand, D. J.\ 2005, \aj, 130, 586

\noindent Ueno, M. 2005, Ph.D. Thesis, Kyoto University

\noindent  Urquhart, J. S., Hoare, M. G., Purcell, C. R., et al.\ 2009, \aap, 501, 539


\clearpage

\begin{table}[]
\begin{center}
\caption[]{
Observation log }
\scriptsize
\vspace{0.4cm}
\begin{tabular}{cccccccc}
\tableline
\multicolumn{1}{c}{Field} & \multicolumn{1}{c}{Obs ID} &\multicolumn{1}{c}{Date} &
\multicolumn{2}{c}{Pointing coordinates} & \multicolumn{1}{c}{ Exposure PN}\tablenotemark{a}  & 
\multicolumn{1}{c}{Exposure MOS1}\tablenotemark{a}  & \multicolumn{1}{c}{Exposure MOS2}\tablenotemark{a}   \\ 
  & & & \multicolumn{2}{c}{R.A.\ and Dec.\ (J2000)} 
& \multicolumn{1}{c}{ks}
& \multicolumn{1}{c}{ks}
& \multicolumn{1}{c}{ks}\\
\tableline
G23.5+0.1 & 0400910101 & 2006-09-16 & 18:33:33.8 & $-$08:25:30.3 &  10.54 & 12.17 &  12.18 \\
G25.5+0.0 & 0400910301 & 2006-10-18 & 18:37:10.0 & $-$06:39:49.7 &  7.53 &   9.17 &   9.17 \\
G25.5+0.0 & 0400910401 & 2006-10-18 & 18:37:43.2 & $-$06:43:38.0 &  7.53 &   9.17 &   9.17 \\
\tableline
\end{tabular}
\label{obslog}
\end{center}
\tablenotetext{a} {Scientific exposures used in the analysis.}
\end{table}

\begin{table}[]
{\scriptsize
\begin{center}
\caption[]{X-ray sources in the G23.5+0.1 field}
\begin{tabular}{ccccccccccc}
\tableline
\tableline
\multicolumn{1}{c}{Source ID\tablenotemark{a}} & \multicolumn{1}{c}{Flux\tablenotemark{b}} &
\multicolumn{1}{c}{$C_{\rm PN}$\tablenotemark{c}} &\multicolumn{1}{c}{$C_{\rm MOS1}$\tablenotemark{c}}   &  \multicolumn{1}{c}{$C_{\rm MOS2}$\tablenotemark{c}}  &
\multicolumn{1}{c}{S/N} & \multicolumn{1}{c}{HR} & \multicolumn{1}{c}{R.A.} &
\multicolumn{1}{c}{Dec.} & \multicolumn{1}{c}{$\sigma_{\rm pos}$\tablenotemark{d}}\\ 
 & \multicolumn{1}{c}{$10^{-14}$ c.g.s. } & \multicolumn{1}{c}{counts/ks}&  \multicolumn{1}{c}{counts/ks} &  \multicolumn{1}{c}{counts/ks} & & &
\multicolumn{1}{c}{deg} & \multicolumn{1}{c}{deg} & \multicolumn{1}{c}{arcsec} \\
\tableline
1 & 21.9$\pm$0.7   &   72.4$\pm$2.9    &  19.3$\pm$1.3  & 18.1$\pm$1.4   & 31.4   & $-0.83\pm0.08$                    & 278.438633 & $-$8.307803 & 1.03  \\
2 & 24.8$\pm$2.3   &  10.1$\pm$1.3     & 4.9$\pm$0.8    & 4.0$\pm$0.7     & 11.3     & $1.0_{-0.10}$                    & 278.324492 & $-$8.420513 & 1.20  \\
3 & 44.0$\pm$6.4   &   4.1$\pm$1.3\tablenotemark{e}       &3.9$\pm$0.7    & 3.6$\pm$ 0.6        & 7.3      & $1.0_{-0.12}$                       & 278.498733 & $-$8.373778 & 1.46 \\
4 & 4.0$\pm$0.8     &  5.0$\pm$1.4\tablenotemark{e}        & 4.3$\pm$0.8      & 3.0$\pm$0.5        & 7.3    & $-0.19\pm0.57$                      & 278.279851 & $-$8.252994 & 1.06  \\
5 & 9.2$\pm$1.8     &  3.3$\pm$1.1\tablenotemark{e}         &  2.8$\pm$0.5      &  2.3$\pm$0.4       & 6.5        & $1.0_{-0.18}$                       & 278.344773 & $-$8.402801 & 1.70 \\
6 & 7.0$\pm$1.1     &  3.2$\pm$1.1         & 2.4$\pm$0.5      & 2.1$\pm$0.4        & 6.0       & $1.0_{-0.26}$                           & 278.340301 & $-$8.509183 & 1.52  \\
7 & 1.9$\pm$0.4         &  3.0$\pm$0.9        & ...                     & 1.8$\pm$0.4        & 5.0        & $-0.05\pm0.75$                       & 278.336846 & $-$8.566790 & 1.96 \\
\tableline
\end{tabular}
\end{center}
\tablecomments{ Properties of X-ray sources found with  the {\em edetect\_chain} automatic detection procedure. The uncertainties correspond to 68\% confidence interval. Since HR cannot exceed 1 by definition, we quote only the lower bound (at
68\%
confidence) in the cases when the formal upper bound is $>1$.
}
\tablenotetext{a} {Source ID number used throughout the paper.}
  \tablenotetext{b} {Observed flux in 0.2--12 keV in units of $10^{-14}$  erg s$^{-1}$ cm$^{-2}$, estimated using ECFs. The uncertainties do not include systematic error which can be significant for strongly absorbed sources  (see \S3).  }
    \tablenotetext{c} {Observed, background-subtracted EPIC count rate for PN, MOS1, and MOS2 in 0.2--12 keV.}
     \tablenotetext{d} {Position uncertainty (see \S 4.2).} 
     \tablenotetext{e}{In PN, the source partly falls within the chip gap which reduces the count rate. }   }
\label{sourceList1}
 \end{table}

\begin{table}[]
 {\scriptsize
\begin{center}
\caption[]{X-ray sources in the G25.5+0.0 field}
\begin{tabular}{ccccccccccc} 
\tableline
\tableline
\multicolumn{1}{c} {Source ID\tablenotemark{a}} & \multicolumn{1}{c}{Flux\tablenotemark{b}} &
\multicolumn{1}{c}{$C_{\rm PN}$\tablenotemark{c}} &\multicolumn{1}{c}{$C_{\rm MOS1}$\tablenotemark{c}} &  \multicolumn{1}{c}{$C_{\rm MOS2}$\tablenotemark{c}} &
\multicolumn{1}{c}{S/N} & \multicolumn{1}{c}{HR} & \multicolumn{1}{c}{R.A.} &
\multicolumn{1}{c}{Dec.} & \multicolumn{1}{c}{$\sigma_{\rm pos}$\tablenotemark{d}}\\ 
& \multicolumn{1}{c}{10$^{-14}$ c.g.s. } & \multicolumn{1}{c}{counts/ks}&  \multicolumn{1}{c}{counts/ks} &  \multicolumn{1}{c}{counts/ks} & & &
\multicolumn{1}{c}{deg} & \multicolumn{1}{c}{deg} & \multicolumn{1}{c}{arcsec} \\
\tableline
1 & 951$\pm$41                &  ...                        &  ...                        &  70.0$\pm$3.8 & 18.4 & $1.0_{-0.06}$ & 279.513756 & $-$6.925462 & 1.07 \\
2 & 11.5$\pm$0.8              &  24.3$\pm$1.3 & 10.4$\pm$1.1    &  9.0$\pm$1.0 & 21.6 & $-0.70\pm0.15$ & 279.495002 & $-$6.805702 &  1.12 \\
3 & 79.5$\pm$6.7              &   17.6$\pm$1.6 & ...                         &   6.5$\pm$0.9 & 13.1 & $1.0_{-0.08}$ & 279.499546 & $-$6.822605 &  1.18 \\
4$^{*}$ & 30.6$\pm$2.8   &   7.5$\pm$0.8   &  2.7$\pm$0.4     &   2.9$\pm$0.5 & 12.7 & $1.0_{-0.13}$ & 279.460173 & $-$6.815297 & 1.21 \\
5$^{*}$ & 9.9$\pm$0.6     &  9.2$\pm$0.9    &  5.7$\pm$0.9      &   3.9$\pm$0.5 & 14.2 &    $0.52\pm0.30$ & 279.262722 & $-$6.823484 &  1.24 \\
6 & 90.8$\pm5.4$              &    30.4$\pm$2.1&  8.6$\pm$1.0     &  9.8$\pm$1.1 & 19.0 & $0.86\pm0.12$ & 279.563358 & $-$6.799838 &  1.30 \\
7$^{*}$ & 8.2$\pm0.4$     &   22.9$\pm$1.2 & 8.1$\pm$1.0      &   4.6$\pm$0.5 & 20.6 & $-0.87\pm0.10$ & 279.298147 & $-$6.554131 & 1.32 \\
8$^{*}$ & 20.7$\pm$2.0   &   5.8$\pm$0.7   &  1.0$\pm$0.3     &  2.1$\pm$0.4 & 9.9 & $1.0_{-0.10}$ & 279.424316 & $-$6.775722 &  1.32 \\
9 & 46.9$\pm$5.9              &    7.2$\pm$1.1  &  ...                         &  3.6$\pm$0.7 & 13.1 & $1.0_{-0.13}$ & 279.464660 & $-$6.895446 &  1.36 \\
10$^{*}$ & 4.3$\pm$0.5   &   5.8$\pm$0.7   &  1.2$\pm$0.3     &  1.8$\pm$0.4 & 9.8 & $0.67\pm0.32$ & 279.404399 & $-$6.735200 &  1.45 \\
11 & 37.8$\pm$4.3            &   5.9$\pm$1.0   &  2.0$\pm$0.4     &  1.5$\pm$0.4 & 7.9 & $1.0_{-0.11}$ & 279.500926 & $-$6.622271 &  1.55 \\
12$^{*}$ & 1.6$\pm$0.2   &    5.8$\pm$0.7  &   1.3$\pm$0.4    &  1.4$\pm$0.4 & 9.5 & $-0.36\pm33$ & 279.388669 & $-$6.667137 &  1.59 \\
13 & 23.8$\pm$4.1            &  3.7$\pm$0.9    &   1.5$\pm$0.5    & 1.5$\pm$0.5 & 5.8 & $1.0_{-0.51}$ & 279.579818 & $-$6.729061 &  1.61 \\
14 & 4.5$\pm$0.7              &  5.5$\pm$0.7    &  1.3$\pm$0.5     &  1.0$\pm$0.3 & 8.3 & $0.69\pm0.30$ & 279.282215 & $-$6.858810 & 1.74  \\
15 & 115$\pm$15              &   17.3$\pm$1.6 & 7.0$\pm$0.9      &  5.5$\pm$0.6 &  15.1 & $1.0_{-0.13}$ & 279.339761 & $-$6.874498 &  2.51 \\
16$^{*}$ & 21.5$\pm$2.7 &   4.6$\pm$0.7   &  2.3$\pm$0.4     &   2.3$\pm$0.4 & 10.2 & $1.0_{-0.14}$ & 279.371524 & $-$6.614574 &  1.87 \\
17 & 8.4$\pm$1.7               &   6.1$\pm$1.0   &  2.1$\pm$0.4    &   1.4$\pm$0.4 & 8.2 & $0.87\pm0.20$ & 279.451552 & $-$6.653508 & 1.63  \\
18 & 6.5$\pm$1.0               &    6.0$\pm$1.0  &  2.4$\pm$0.6    &   1.4$\pm$0.5 & 7.5 & $0.67\pm0.31$ & 279.202677 & $-$6.540072 & 1.66 \\
19 & 2.4$\pm$0.4               & 2.6$\pm$0.6     & 1.5$\pm$0.4     &  1.4$\pm$0.4 & 7.1 & $-0.04\pm0.37$ & 279.336448 & $-$6.538502 & 1.74  \\
20 & 9.9$\pm$2.1               &  2.8$\pm$0.8    &  0.9$\pm$0.4    &  1.4$\pm$0.5 & 4.8 & $1.0_{-0.26}$ & 279.145257 & $-$6.728290 & 1.90 \\
21 & 6.2$\pm$1.2               &  4.0$\pm$0.6    & 1.0$\pm$0.3     &  1.1$\pm$0.3 & 7.7 & $1.0_{-0.36}$ & 279.393015 & $-$6.683063 & 1.76 \\
22 &1.0$\pm$0.2                &  3.5$\pm$0.6    &  1.1$\pm$0.3    &   0.9$\pm$0.3 & 7.1 & $-0.14\pm0.73$ & 279.359209 & $-$6.704057 & 1.67 \\
23 & 1.4$\pm$0.3               &  2.9$\pm$0.8    &   1.0$\pm$0.3   &   0.7$\pm$0.3 & 5.0 & $0.24\pm0.71$ & 279.232527 & $-$6.710708 & 1.73 \\
\tableline
\end{tabular}
\end{center}
\tablecomments{  Properties of X-ray sources from the {\em edetect\_chain} automatic detection procedure. Asterisks denote the sources detected in both observations of  the G25.5+0.0 field. The uncertainties correspond to 68\% confidence interval. Since HR cannot exceed 1 by definition, we quote only the lower bound (at
68\%
confidence) in the cases when the formal upper bound is $>1$.
}
\tablenotetext{a} {Source ID number used throughout the paper. }
  \tablenotetext{b} {Observed flux in 0.2--12 keV in units of $10^{-14}$  erg s$^{-1}$ cm$^{-2}$, estimated using ECFs. The uncertainties do not include systematic error which can be significant for strongly absorbed sources  (see \S3).  }
   \tablenotetext{c} {Observed, background-subtracted EPIC count rates for PN, MOS1, and MOS2 in 0.2--12 keV (averaged if the source is detected in both fields.}
    \tablenotetext{d} {Position uncertainty (see \S 4.2).}   }
\label{sourceList2}
\end{table}

\begin{table}[]
\caption{Absorbed {\em mekal} model fits to the  spectra of sufficiently bright sources}
 {\footnotesize
\begin{center}
\begin{tabular}{ccccccc}
\tableline\tableline
Field (Source ID) &$N_H$ &$kT$ & Abundance\tablenotemark{a} & \multicolumn{1}{c}{Normalization\tablenotemark{b}} & \multicolumn{1}{c}{Flux\tablenotemark{c}} & \multicolumn{1}{c}{$\chi_\nu^{2}$}/d.o.f. \\ 
 & 10$^{21}$ cm$^{-2}$ & keV & \% Solar & &10$^{-13}$ erg cm$^{-2}$ s$^{-1}$ &  \\ 
\tableline
G23.5+0.1 (Source 1) & $1.3\pm0.2$ & $0.59\pm0.24$ & 10  & $6.0^{+0.6}_{-0.5}\times10^{-4}$ & 3.5$\pm$0.3 & 1.20/71 \\
G25.5+0.0 (Source 6) & $28.3\pm4.2$ & $2.20^{+0.37}_{-0.26}$ & 110 & $8.4^{+1.8}_{-1.6}\times10^{-4}$ & 9.6$\pm$1.9 & 1.05/52 \\
G25.5+0.0 (Source 7) & $3.8^{+1.7}_{-0.5}$ & $0.28\pm0.07$ & 330 & $7.5^{+3.4}_{-0.7}\times10^{-4}$ & 4.6$\pm$2.5 & 0.99/21 \\
\tableline
\end{tabular}
\end{center} 
\tablecomments{  The uncertainties are given at the 68\% confidence
level for a single interesting parameter.
}
\tablenotetext{a} {Abundance is the number of metal nuclei per Hydrogen nucleus relative to the solar value. Helium abundance is fixed at the cosmic value (9.77\%).  (See http://heasarc.gsfc.nasa.gov/docs/xanadu/xspec/manual/XSmodelMeka.html for details).}
\tablenotetext{b} {
 Normalization of the {\em mekal} model defined at  http://heasarc.gsfc.nasa.gov/docs/xanadu/xspec/manual/XSmodelMeka.html. }
\tablenotetext{c} {Unabsorbed fluxes given for the same energy bands as the absorbed fluxes in the text, i.e., 0.3--5 keV,  1--8 keV, and 0.3--2 keV (top to bottom, respectively).}}
\label{specMEKAL}
\end{table}

\begin{table}[]
\caption{Absorbed PL model fits to the spectra of sufficiently bright sources}
 {\footnotesize
\begin{center}
\begin{tabular}{cccccc}
\tableline\tableline
Field (Source ID) &$N_H$ & $\Gamma$ & Normalization\tablenotemark{a} &Flux\tablenotemark{b} & $\chi_\nu^{2}$/d.o.f. \\ 
 &  10$^{22}$ cm$^{-2}$ & & &10$^{-13}$ erg cm$^{-2}$ s$^{-1}$ &  \\ 
\tableline
G23.5+0.1 (Source 8) & $3.9\pm1.9$ & $2.3\pm0.8$ & $1.8^{+9.0}_{-1.8}\times10^{-4}$ & 5.6$\pm$3.0 & 1.31/48 \\
G25.5+0.0 (Source  1) & $5.2^{+1.0}_{-0.8}$ & $1.2^{+0.3}_{-0.1}$ & $1.1^{+0.7}_{-0.4}\times10^{-3}$ & 124$\pm$20 & 0.70/41  \\
G25.5+0.0 (Source 6) & $2.9^{+0.6}_{-0.5}$ & $2.8^{+0.4}_{-0.3}$ & $5.8^{+4.1}_{-2.2}\times10^{-4}$ & 9.5$\pm$1.8 & 1.31/52 \\
G25.5+0.0 (Source 15) & $6.7^{+3.8}_{-2.2}$ & $2.1^{+0.9}_{-0.6}$ & $2.8^{+10.5}_{-2.8}\times10^{-4}$ & 7.4$\pm$2.3 & 0.95/45 \\
\tableline
\end{tabular}
\end{center} 
\tablecomments{  The uncertainties are given at the 68\% confidence
level for a single interesting parameter.
}
\tablenotetext{a} {Spectral flux density at 1 keV, in photons s$^{-1}$ cm$^{-2}$ keV$^{-1}$.}
\tablenotetext{b} {Unabsorbed fluxes given for the same energy bands as the absorbed fluxes in the text, i.e. 1--10 keV,  1--11 keV,  1--8 keV, and 2--10 keV (top to bottom, respectively).}}
\label{specPL}
\end{table}

{
\begin{deluxetable}{cccccccccccrcccr}
\rotate
\voffset=4cm
\hoffset=-2cm 
\tabletypesize{\tiny}
\setlength{\tabcolsep}{0.05in} 
\tablecaption{Candidate optical and infrared counterparts and the proposed
  classification of the X-ray
  sources  in the G23.5+0.1 field.  } 
  \tablehead{
  \colhead{Source } &   
 \colhead{Offset\tablenotemark{a}} &  \colhead{Prob.\tablenotemark{b}} &  \colhead{$J$\tablenotemark{c}} &
 \colhead{$H$\tablenotemark{c}} & \colhead{$K$\tablenotemark{c}} &  \colhead{Offset\tablenotemark{d}} &  \colhead{Prob.\tablenotemark{e}} &
 \colhead{$V$\tablenotemark{f}} &  \colhead{$I$\tablenotemark{f}} &  \colhead{$\log(\frac{f_{X} }{ f_{V}})$\tablenotemark{g}} & \colhead{Offset\tablenotemark{h}} & \colhead{4.5mag\tablenotemark{i}} &  \colhead{8.0mag\tablenotemark{i}} &   \colhead{Class\tablenotemark{j}} \\
 &  \colhead{arcsec} & & & & & \colhead{arcsec} & & & & & \colhead{arcsec} & & &   \colhead{}
  }
   \tablecolumns{15}
\startdata
1 & $1.3\pm1.1$ & 0.92 & 8.5$\pm$0.03 & 8.2$\pm$0.06 & 8.0$\pm$0.02 &      $1.3\pm1.1$ & 0.96 & 10.1 & 9.2 & -3.1 & $1.4\pm1.1$ & 8.03$\pm$0.04 & 7.93$\pm$0.03 & F8-star \\ 
2 & $0.9\pm1.2$ & 0.96 & 15.5$\pm$0.09 & 14.2$\pm$0.10 & 13.7$\pm$0.08 &     ...   & ... & ...& ...   & $>0.1$    & $0.8\pm1.2$ & 13.25$\pm$0.23 &... &CV?\\ 
3 & ... & ... & ...  & ...  & ...  & ... & ... & ... & ... & $>0.1$ & $3.9\pm1.5$ & 13.34$\pm$0.23 & ... & AGN/CV/NS \\ 
4 &    
$0.5\pm1.1$ & 0.99 & 12.2$\pm$0.03 & 11.4$\pm$0.03 & 11.1$\pm$0.03 &   $0.4\pm1.1$ & 0.99 & 16.9 & 14.2 & -1.1 & $0.7\pm1.1$ & 10.81$\pm$0.06 & 10.77$\pm$0.10 & M/K-dwarf \\
5 &                                   
$1.4\pm1.7$ & 0.91 & 17.4$\pm$0.00 & 15.1$\pm$0.00 & 13.8$\pm$0.09 &  ...      & ... & ... &  ...    & $>$-0.3  & $1.4\pm1.7$ & 12.76$\pm$0.11 &... & AGN/PMS-star/CV\\ 
6 & ...   &    ...      &              ...                &          ...                   &          ...                   &    ...    & ... & ... & ...      & $>$-0.2 & ...       &       ...                         & ... & AGN/CV/NS?  \\
7 &                  
$1.5\pm2.9$ & 0.90 & 7.2$\pm$0.02 & 7.0$\pm$0.05 & 7.0$\pm$0.02 &      $1.4\pm2.0$ & 0.96 & 8.1 & 7.6 & -5.2& $1.3\pm2.0$ & 6.95$\pm$0.05 & 6.91$\pm$0.03 & A5-star\\  
\enddata
\tablecomments{  The uncertainties are given at the 68\% confidence}
\tablenotetext{a}{ Offset between the X-ray  source and its 2MASS counterpart.}\tablenotetext{b}{Probability of the 2MASS source being a true counterpart to the X-ray source. }\tablenotetext{c}{Magnitudes from the 2MASS All-Sky Catalog of Point Sources (Skrutskie et al. 2006).}\tablenotetext{d}{Offset between the X-ray  source and its USNO-B1.0 counterpart }\tablenotetext{e}{Probability of the USNO-B1.0  source being a true counterpart to the X-ray source. }\tablenotetext{f}{Magnitudes from the USNO-B1.0  catalog (Monet et al.\  (2003)).}\tablenotetext{g}{ Optical-to-X-ray flux ratio (see \S5).} \tablenotetext{h}{Offset between the X-ray  source and its GLIMPSE counterpart. } 
\tablenotetext{i}{Magnitudes from the GLIMPSE  catalog (http://irsa.ipac.caltech.edu/data/SPITZER/GLIMPSE/).} \tablenotetext{j}{Probable classification of the sources.}
\end{deluxetable}
}

\begin{deluxetable}{lrrrrrrrrrrrrrr}
\rotate
\tabletypesize{\tiny}
\setlength{\tabcolsep}{0.02in} 
\tablecaption{Candidate optical and infrared counterparts and the proposed
  classification of the X-ray
  sources  in the G25.5+0.0 field.  } 
  \tablehead{
  \colhead{Source } &   
 \colhead{Offset\tablenotemark{a}} &  \colhead{Prob.\tablenotemark{b}} &  \colhead{$J$\tablenotemark{c}} &
 \colhead{$H$\tablenotemark{c}} & \colhead{$K$\tablenotemark{c}} &  \colhead{Offset\tablenotemark{d}} &  \colhead{Prob.\tablenotemark{e}} &
 \colhead{$V$\tablenotemark{f}} &  \colhead{$I$\tablenotemark{f}} &  \colhead{$\log(\frac{f_{X} }{ f_{V}})$\tablenotemark{g}} & \colhead{Offset\tablenotemark{h}} & \colhead{4.5mag\tablenotemark{i}} &  \colhead{8.0mag\tablenotemark{i}} &   \colhead{Class\tablenotemark{j}} \\
 &  \colhead{arcsec} & & & & & \colhead{arcsec} & & & & & \colhead{arcsec} & & &   \colhead{}
  }
   \tablecolumns{15}
\startdata
1  & ... &   ...      &         ...                     &         ...                    &          ...                   &      ...                 &      ...                 &    ...          &  ...          &        $>1.8$                & ... &  ...       & ...  & PSR      \\       
2 &                       
$1.1\pm1.2$ & 0.96 & 6.9$\pm$0.02 & 6.5$\pm$0.02 & 6.4$\pm$0.02 &     $0.8\pm1.2$ & 0.99 & 8.4 & 7.3 & $-4.1$ & $0.5\pm1.2$ & ... & 6.37$\pm$0.03   & G5-star\\ 
3 & ... & ... & ... & ... & ...  & ... & ... & ...  & ...  & $>0.4$ & $2.9\pm1.2$ & 13.69$\pm$0.22 & ... & AGN/CV/PMS-star?  \\ 
4$^{*}$ &     
$3.6\pm1.2$ & 0.64 & 15.0$\pm$0.06 & 14.4$\pm$0.07 & 12.7$\pm$0.00 &  $3.4\pm1.2$ & 0.80 & 18.9 & 16.8 & $>$-0.5 & $3.2\pm1.2$ & 13.23$\pm$0.20 & ... &CV? \\ 
5$^{*}$ &  
$2.1\pm1.3$ & 0.86 & 9.4$\pm$0.02 & 8.4$\pm$0.02 & 8.0$\pm$0.02 &     $1.6\pm1.3$ & 0.96 & ... & 12.0 & $>0.6$ & $1.4\pm1.3$ & 7.85$\pm$0.04 & 7.72$\pm$0.03  & M-dwarf\\
6 &                   
$1.3\pm1.3$ & 0.95 & 10.0$\pm$0.02 & 8.7$\pm$0.02 & 7.8$\pm$0.02 &    $0.8\pm1.3$ & 0.99 & ... & 12.9 & $>1.3$ & $0.0\pm1.3$ & ... & $\sim 0.2\tablenotemark{a}$ & W42 core\\ 
7$^{*}$ &                 
 $2.2\pm1.3$ & 0.84 & 8.8$\pm$0.02 & 8.6$\pm$0.02 & 8.5$\pm$0.02 &     $2.3\pm1.3$ & 0.91 & 9.8 & 9.1 & $-3.6$ & $1.3\pm1.3$ & 8.53$\pm$0.06 & 8.48$\pm$0.04 & F8-star  \\ 
8$^{*}$ &              
$0.3\pm1.3$ & 0.99 & 15.7$\pm$0.08 & 15.0$\pm$0.00 & 14.6$\pm$0.00 & $ 0.5\pm1.3$ & 0.99 & ... & 17.4 & $>$-0.6  & $2.6\pm1.3$ & 13.06$\pm$0.20 & ... & AGN?  \\ 
9  & ... &  ...       &       ...                       &       ...                      &         ...                    &        ...               & ...   &   ...        &  ...      &     $>0.1$      &      ...                        & ...  &  ...  &  AGN/MQSO    \\       
10$^{*}$ &      
$1.5\pm1.5$ & 0.92 & 12.1$\pm$0.04 & 11.1$\pm$0.04 & 10.6$\pm$0.06 &  $1.4\pm1.5$ & 0.96 & 17.3 & 13.3 & $-1.4$ & $2.2\pm1.5$ & 10.51$\pm$0.06 & 11.02$\pm$0.15 & K/M-dwarf\\
11 &            
$2.9\pm1.6$ & 0.74 & 15.7$\pm$0.10 & 13.2$\pm$0.00 & 12.2$\pm$0.00 & ...  & ... & ... &  ...  & $>$-0.4   & $2.7\pm1.6$ & 12.08$\pm$0.12 &... & AGN/CV/PMS-star? \\
12$^{*}$ &   
$1.5\pm1.6$ & 0.92 & 12.3$\pm$0.03 & 11.7$\pm$0.02 & 11.5$\pm$0.02 &  $1.3\pm1.6$ & 0.97 & 15.6 & 13.2 & $-2.1$ & $1.3\pm1.6$ & 11.20$\pm$0.08 & 11.15$\pm$0.14  &K-dwarf\\  
13  & ... &  ...       &         ...                     &         ...                    &            ...                 &          ...             &    ...       & ...  &    ...        &     $>$-0.2           &   ...         &         ...                        & ...   &  AGN/CV/NS?    \\       
14 &       
$1.0\pm1.8$ & 0.97 & 13.4$\pm$0.03 & 12.8$\pm$0.04 & 11.6$\pm$0.00 &  $1.1\pm1.8$ & 0.98 & 16.2 & 13.7 &$-1.7$ & $3.2\pm1.8$ & 12.30$\pm$0.09 & ... &M-dwarf\\  
15  & ... &   ...      &         ...                     &        ...                     &          ...                   &       ...                &     ...                  &    ...            &   ...         &   $>0.9$              & ...  &   ...            &  ... & PWN      \\       
16 & ... & ... & ... & ... & ... & ... & ...  & ...  & ...  & $>$-0.1 & $1.5\pm1.9$ & 12.58$\pm$0.12 & 11.01$\pm$0.15 & AGN/PMS-star\\ 
17 & ...  & ...  & ...  & ...  & ...  & ...  & ...  & ... & ... & $>0.2$& $0.0\pm1.7$ & 12.18$\pm$0.11 &... & AGN/PMS-star\\
18 & ... & ...  & ... & ... & ... &  ... & ...  & ...  & ... &$>0.3$& $1.4\pm1.7$ & 12.60$\pm$0.13 &... & AGN/PMS-star\\
19 & ... &... & ...&... &... &... &... &... &... &$>0.2$&...  & ... & ...& INS?\\  
20 & ... &... &... &... & ...&... &... &... &... &$>$-0.3 & $1.2\pm1.9$ & 12.64$\pm$0.11 &... & AGN/PMS-star\\ 
21 &                                   
$2.2\pm1.8$ & 0.85 & 13.4$\pm$0.04 & 12.1$\pm$0.05 & 11.3$\pm$0.05 & ...  & ... & ... &  ...  &  $>0.0$  & $1.6\pm1.8$ & 10.73$\pm$0.15 & 10.25$\pm$0.16 & K/M-dwarf/PMS-star \\ 
22 & ... & ... & ... & ...& ...& ... & ... & ... & ... & $>$-0.2 & $0.7\pm1.7$ & 12.10$\pm$0.11 &... & F/G/K/M-dwarf/PMS-star\\ 
23 &        
$1.7\pm1.8 $ & 0.90 & 13.8$\pm$0.03 & 12.8$\pm$0.04 & 12.6$\pm$0.30 &  $1.4\pm1.8$ & 0.96 & 17.6 & 15.2 & $-1.4$ & $1.9\pm1.8$ & 12.28$\pm$0.10 & ... &K/M-dwarf  \\
\enddata
\tablecomments{  The uncertainties are given at the 68\% confidence}
\tablenotetext{a}{ Offset between the X-ray  source and its 2MASS counterpart.}\tablenotetext{b}{Probability of the 2MASS source being a true counterpart to the X-ray source. }\tablenotetext{c}{Magnitudes from the 2MASS All-Sky Catalog of Point Sources (Skrutskie et al. 2006).}\tablenotetext{d}{Offset between the X-ray  source and its USNO-B1.0 counterpart }\tablenotetext{e}{Probability of the USNO-B1.0  source being a true counterpart to the X-ray source. }\tablenotetext{f}{Magnitudes from the USNO-B1.0  catalog (Monet et al.\  (2003)).}\tablenotetext{g}{ Optical-to-X-ray flux ratio (see \S5).} \tablenotetext{h}{Offset between the X-ray  source and its GLIMPSE counterpart. }  
\tablenotetext{i}{Magnitudes from the GLIMPSE  catalog (http://irsa.ipac.caltech.edu/data/SPITZER/GLIMPSE/).} \tablenotetext{j}{Probable classification of the sources.}
\end{deluxetable}

\clearpage

\begin{figure}[]
 \centering
 \vspace{-0.5cm}
\includegraphics[width=2.2in]{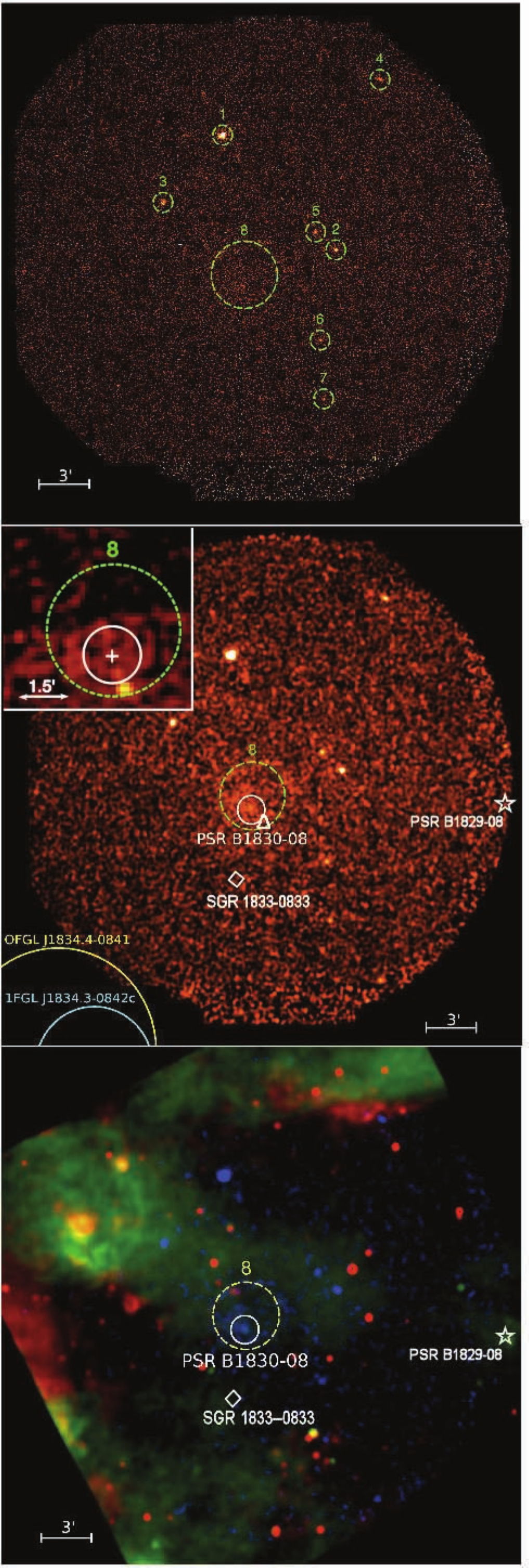}
\caption{{\em Top:} Combined PN+MOS1+MOS2  image  of  the G23.5+0.1 field in the 0.2--10 keV band.  The small green circles mark point-like (or compact) X-ray sources found by the SAS source detection procedure. The large green circle shows the  diffuse Source 8 identified visually (see \S3).
  Source numbering corresponds to that in  Table 2. {\em Middle:} The same image 
  as in the top panel but  smoothed  with the gaussian (FWHM  6$^{\prime\prime}$). The white circle shows the position of PSR B1830--08. Small white triangle shows the position of  CXOU J183340.3--082830  found in the {\sl Chandra} image (see text). The inset shows the {\sl Chandra} ACIS image (discussed in \S) with the same circles as in the EPIC image and the cross at the radio position of the pulsar.   {\em Bottom:}  Multiwavelength image of G23.5+0.1 field (blue:  0.2--10 keV combined EPIC image; green:  20 cm radio image from MAGPIS; red:  {\sl Spitzer} IRAC image  at 8.0 $\mu$m).}
\end{figure}

\begin{figure}[]
 \centering
 \vspace{-0.7cm}
\includegraphics[width=2.1in]{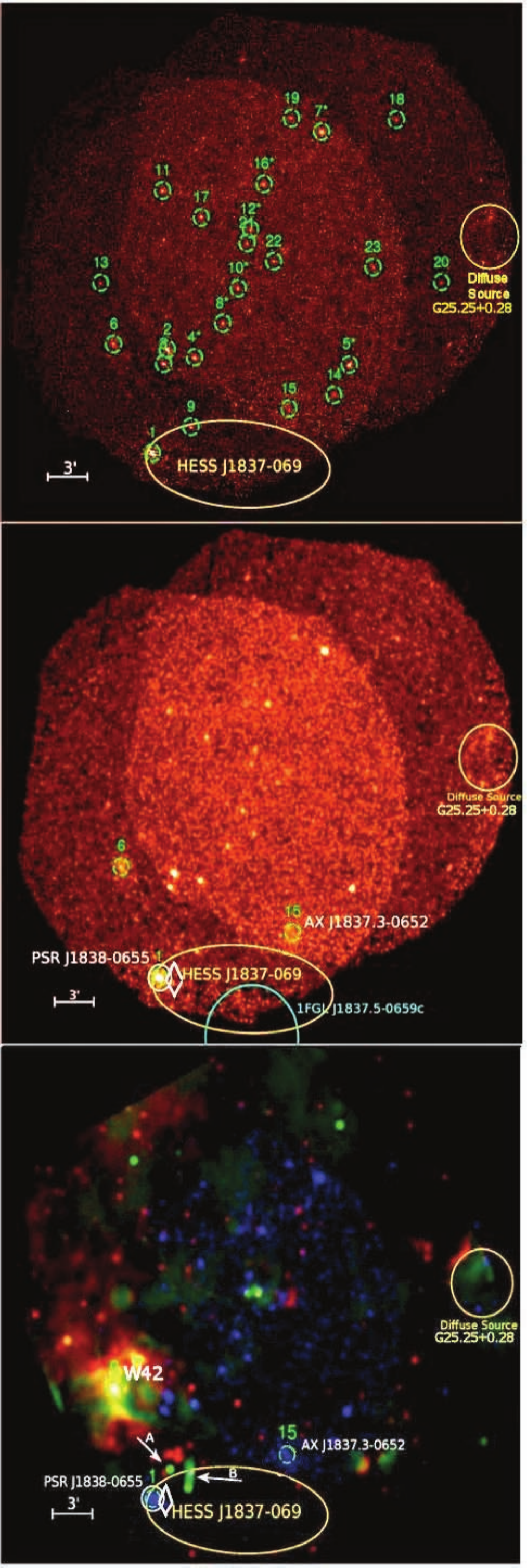}
\caption{{\em Top:} Combined PN+MOS1+MOS2  image  of the G25.5+0.0 field in the 0.2--10 keV band obtained  by 
  merging the data from two {\sl XMM-Newton} observations  of the field (see Table 1).    The small green circles mark point-like  X-ray sources found by the automatic source detection procedure. The yellow ellipse near the western edge of the FOV shows a  diffuse shell-like emission (G25.25+0.28; see \S4.3),  which was not detected automatically.
  The source numbering corresponding to that in Table 3. The yellow elliptical region 
  at the southern  edge of the FOV  shows the 1$\sigma$ extent of the TeV emission from HESS J1837--069. {\em Middle:} The same image 
  as in the top panel but  smoothed  with the gaussian (FWHM  6$^{\prime\prime}$). The white diamond region marks the positions of the {\sl INTEGRAL} source coincident with AX J1838.0--065. 
   The white circle  marks PSR J1838--0655  and its PWN coincident with Source 1.  {\em Bottom:} Multiwavelength  image of G25.5+0.0 field (blue: 0.2--10 keV combined EPIC image; green: 20 cm radio image from MAGPIS; red: {\sl  Spitzer} IRAC image  at 8.0 $\mu$m).}

\end{figure}

\begin{figure}[]

 \centering

\includegraphics[width=3.0in]{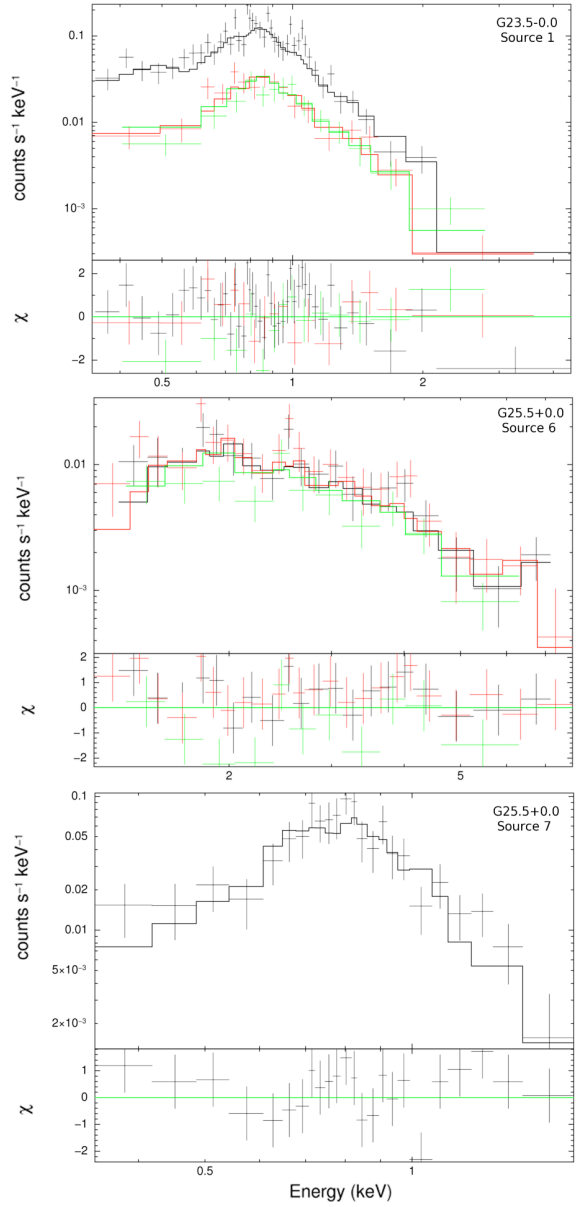}

\caption{ Absorbed {\em mekal} model fits  to the EPIC spectra of Sources 1 (G23.5+0.1),  6 (G25.0+0.0), and 7 (G25.0+0.0).  The best-fit parameters are given in Table 4.
 }
\end{figure}

\begin{figure}[]
 \centering
\includegraphics[width=7.0in]{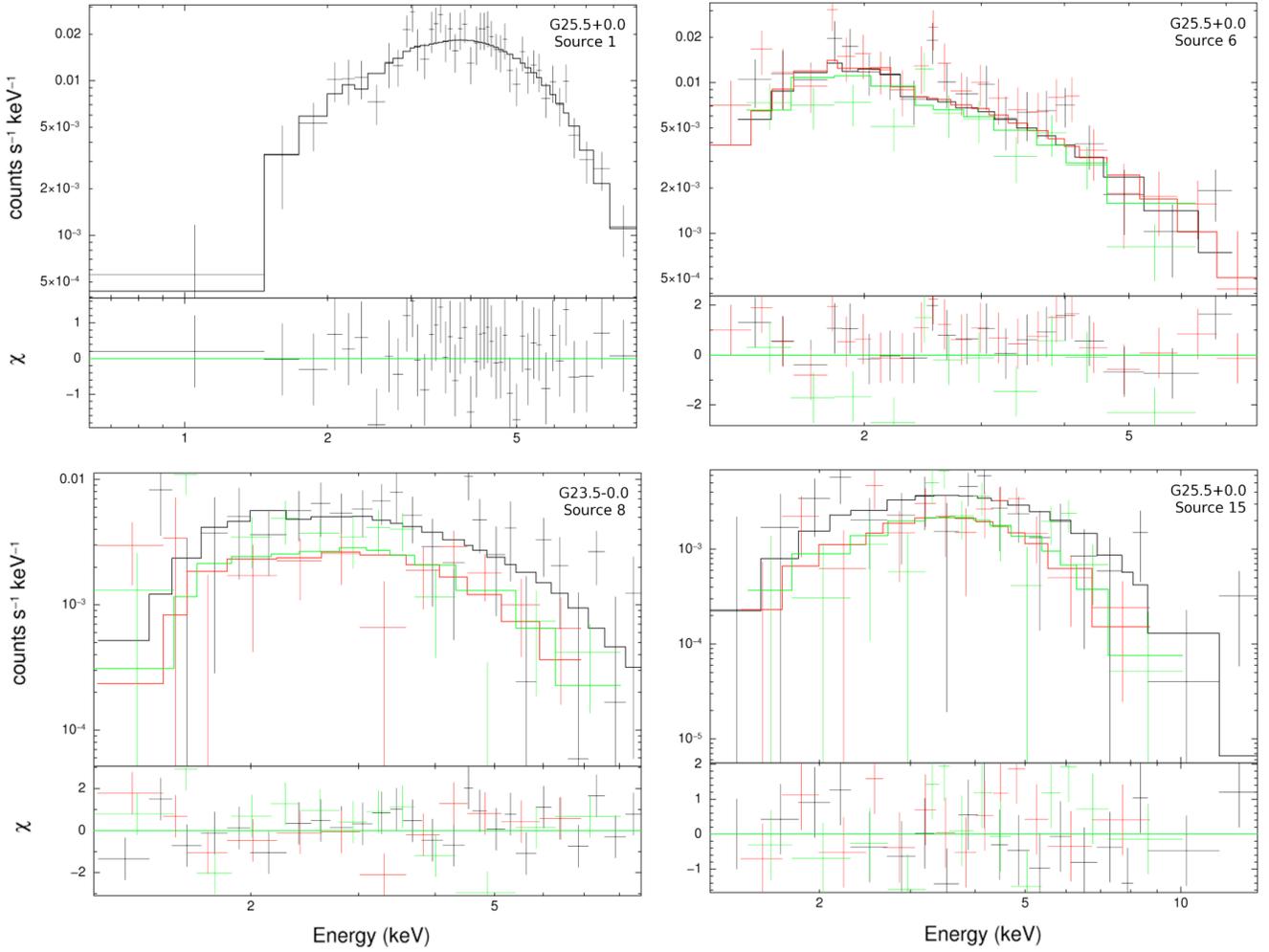}
\caption{ Absorbed PL model fits to the EPIC spectra of Source 8 in G23.5+0.1, and Sources 1, 6, and 15 in G25.5+0.0. The best-fit parameters are given in Table 5.
 }

\end{figure}

\begin{figure}[]

 \centering

\includegraphics[width=7.0in]{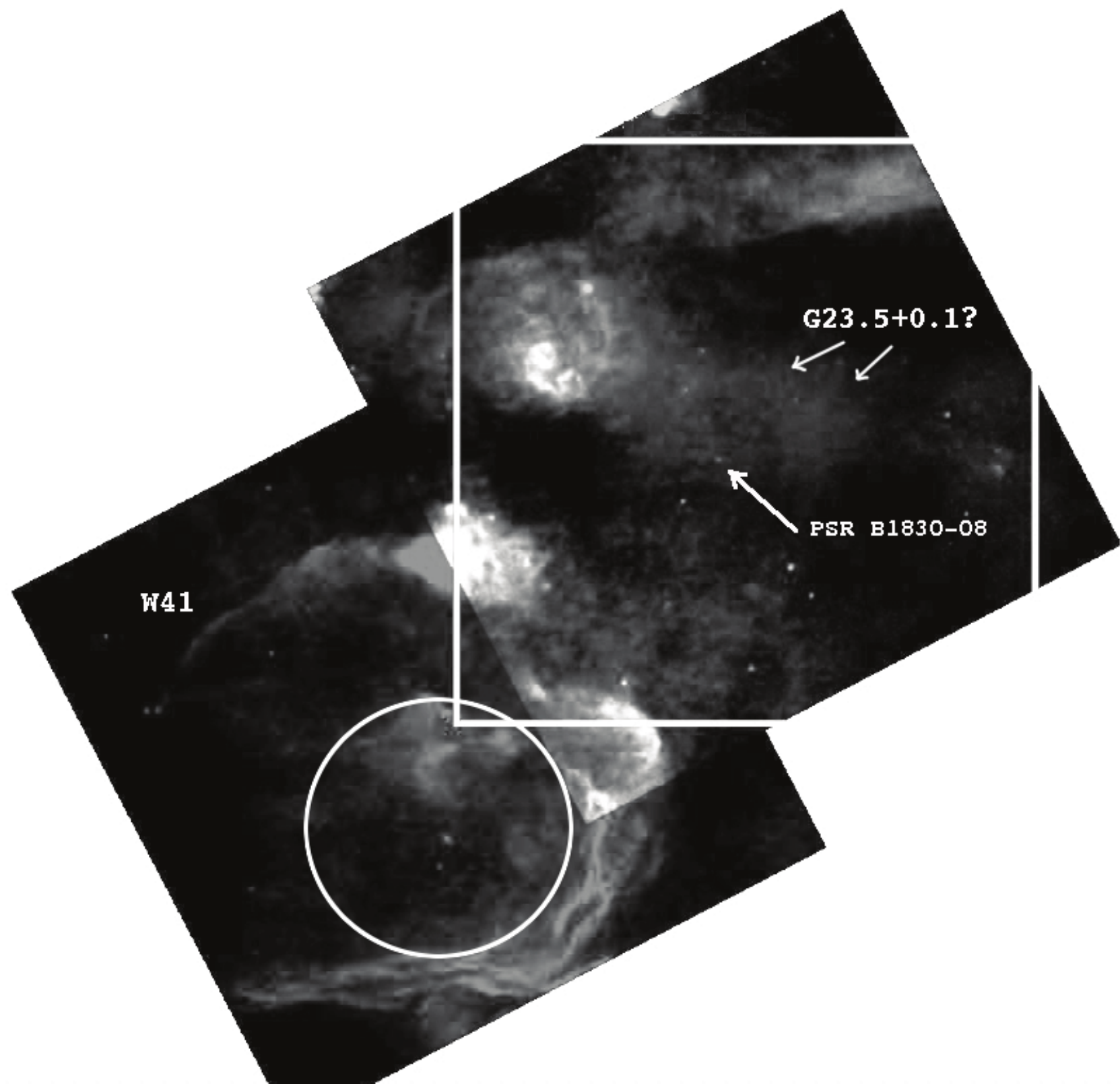}

\caption{MAGPIS  20 cm  composite image (1.4 GHz) showing the G23.5+0.1 and  W41  fields  together. The  $32'\times32'$ white box encloses  the {\sl XMM-Newton} EPIC FOV shown in Figure 1. The shell of SNR W41 is clearly seen.   W41  appears to have no connection to PSR B1830--08 and surrounding it diffuse emission. The $r=7'$ circle shows 1$\sigma$ extent of the TeV source HESS~J1834--087.
}

\end{figure}

\begin{figure}[]

 \centering

\includegraphics[width=7.0in]{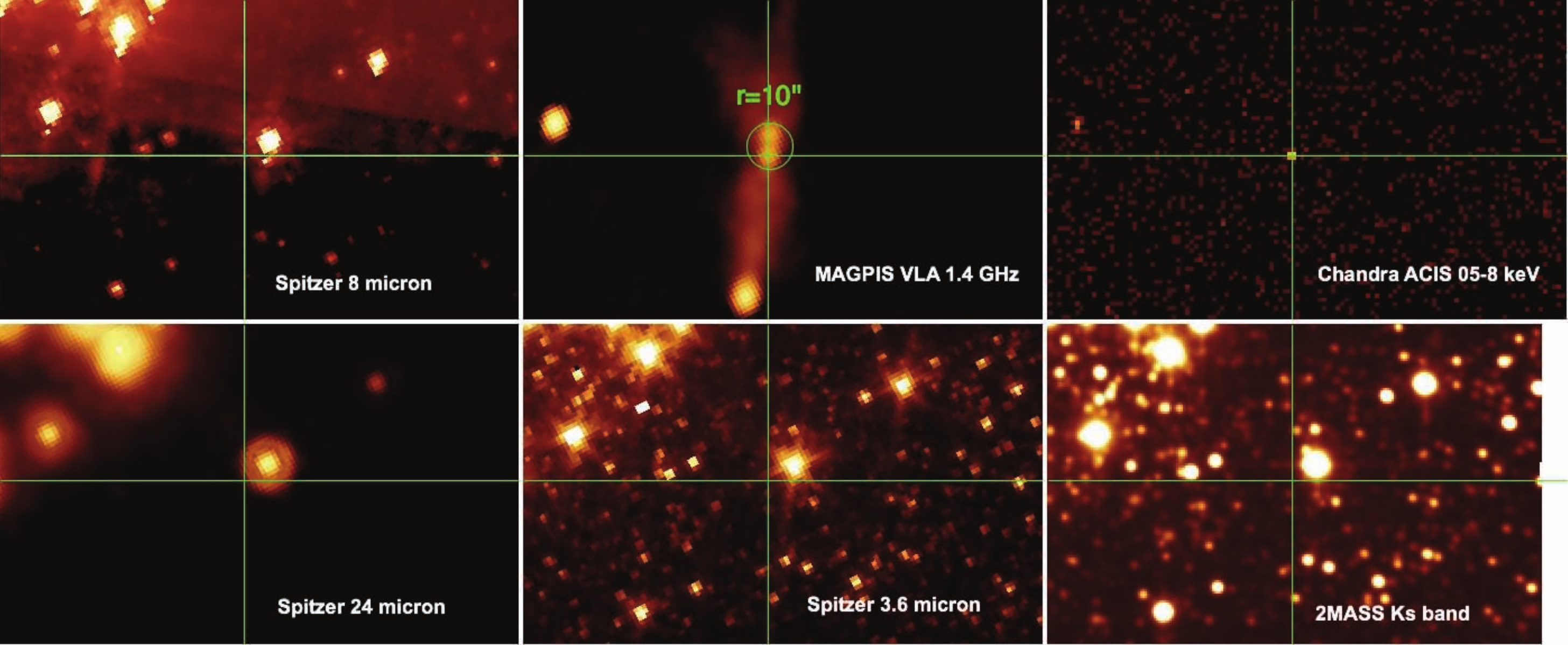}

\caption{ Radio source B ({\sl XMM-Newton} Source 9 in the G25.5+0.0 field) at different wavelengths (see \S4.3). The cross-hair is centered at the position of the {\sl Chandra} source,  CXOU~J183751.62--064355.4,  which is coincident with  Source 9 (see top right panel). }

\end{figure}

\begin{figure}[]

 \centering

\includegraphics[width=3.3in,angle=90]{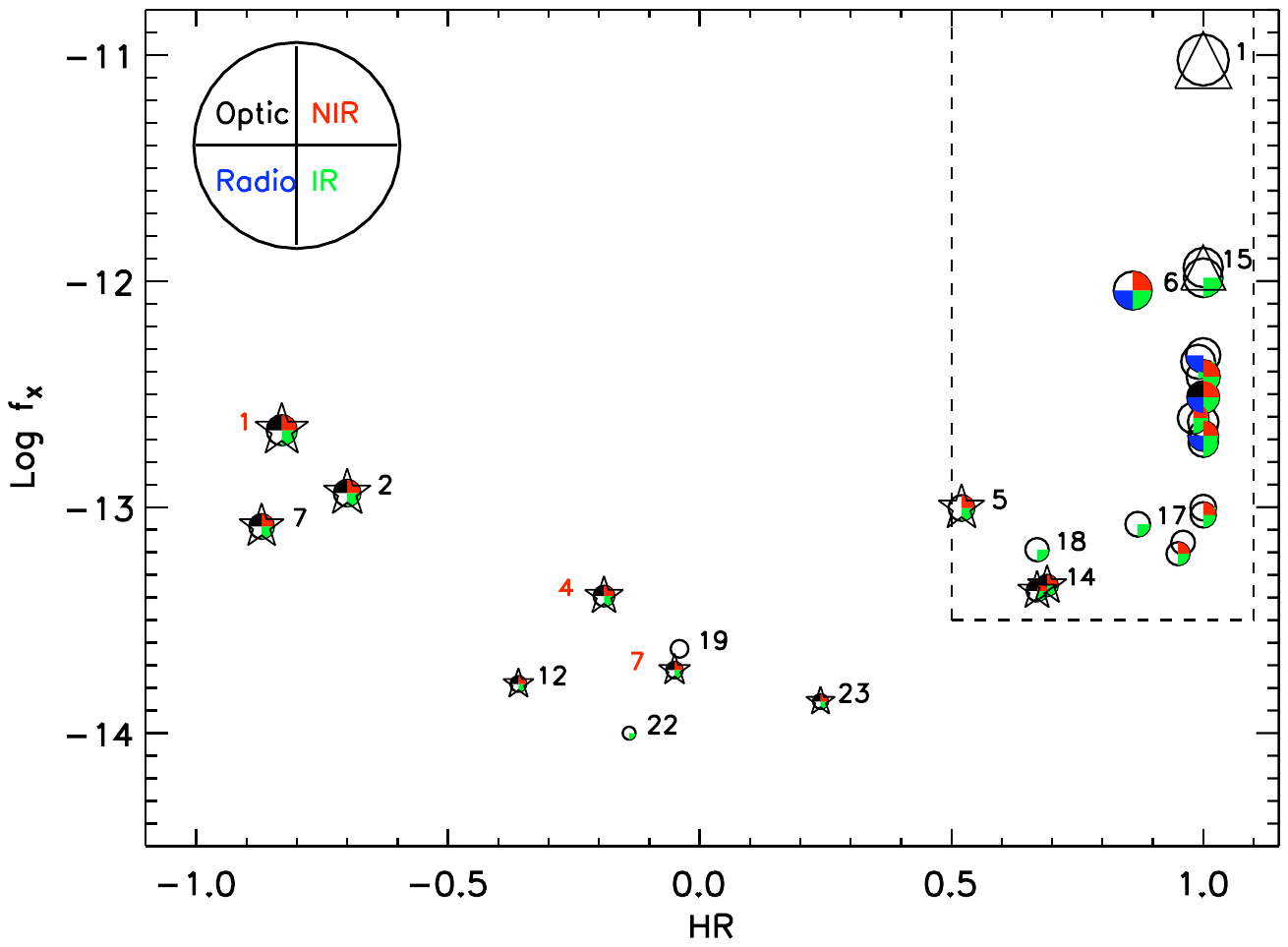}\vspace{-0.5cm}
\includegraphics[width=3.3in,angle=90]{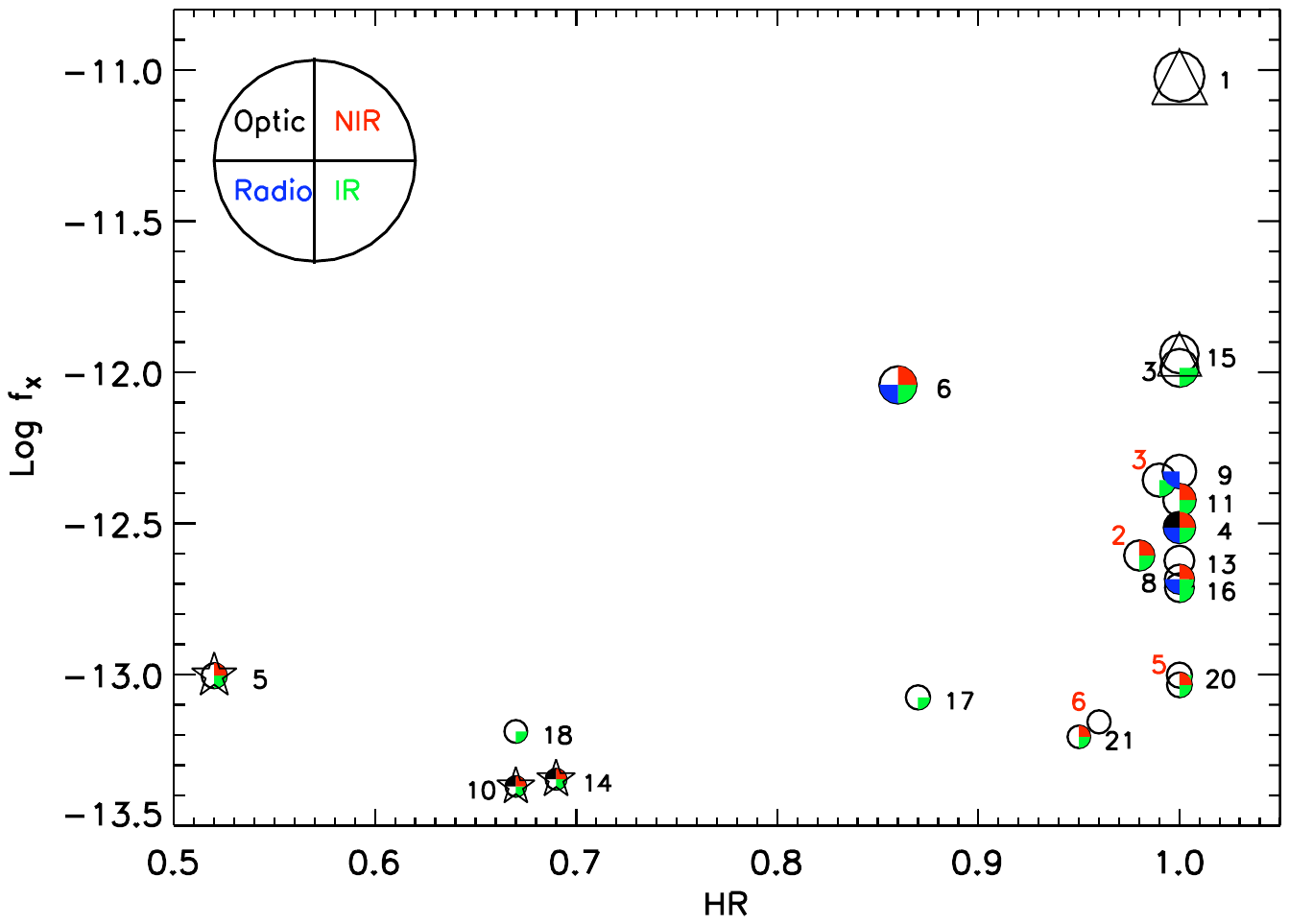}\vspace{-0.5cm}

\caption{ \footnotesize{Logarithm of the observed 0.2--12 keV X-ray flux  ($f_{X}$) versus hardness ratio (HR) for the X-ray sources detected in G23.5+0.1 and G25.5+0.0 fields (denoted by red and black numbers, respectively, 
in accordance with Tables 2 and 3).  The bottom panel is the zoom-in of the upper right corner  
of  the top panel (shown by the dashed lines).  The sizes of  the circles are proportional to $\log f_X$.  The optical, 2MASS, GILMPSE, and radio detections are shown by 
 filling the circle quadrants with different colors, according to the legend shown in the upper left corner. An empty (white) quadrant means nondetection in the corresponding wavelength range. Sources classified as foreground MS stars are additionally marked  by the ``star'' symbol, while those classified as an AGN or PMS-star are are marked by the ``square'' symbol. The young pulsar and the  PWN candidate are marked by the ``triangle'' symbols. Classification of the sources shown by circles only (without additional symbols)  is uncertain (see Tables 6 and 7). }}

\end{figure}

\begin{figure}[]

 \centering

\includegraphics[width=3.3in,angle=90]{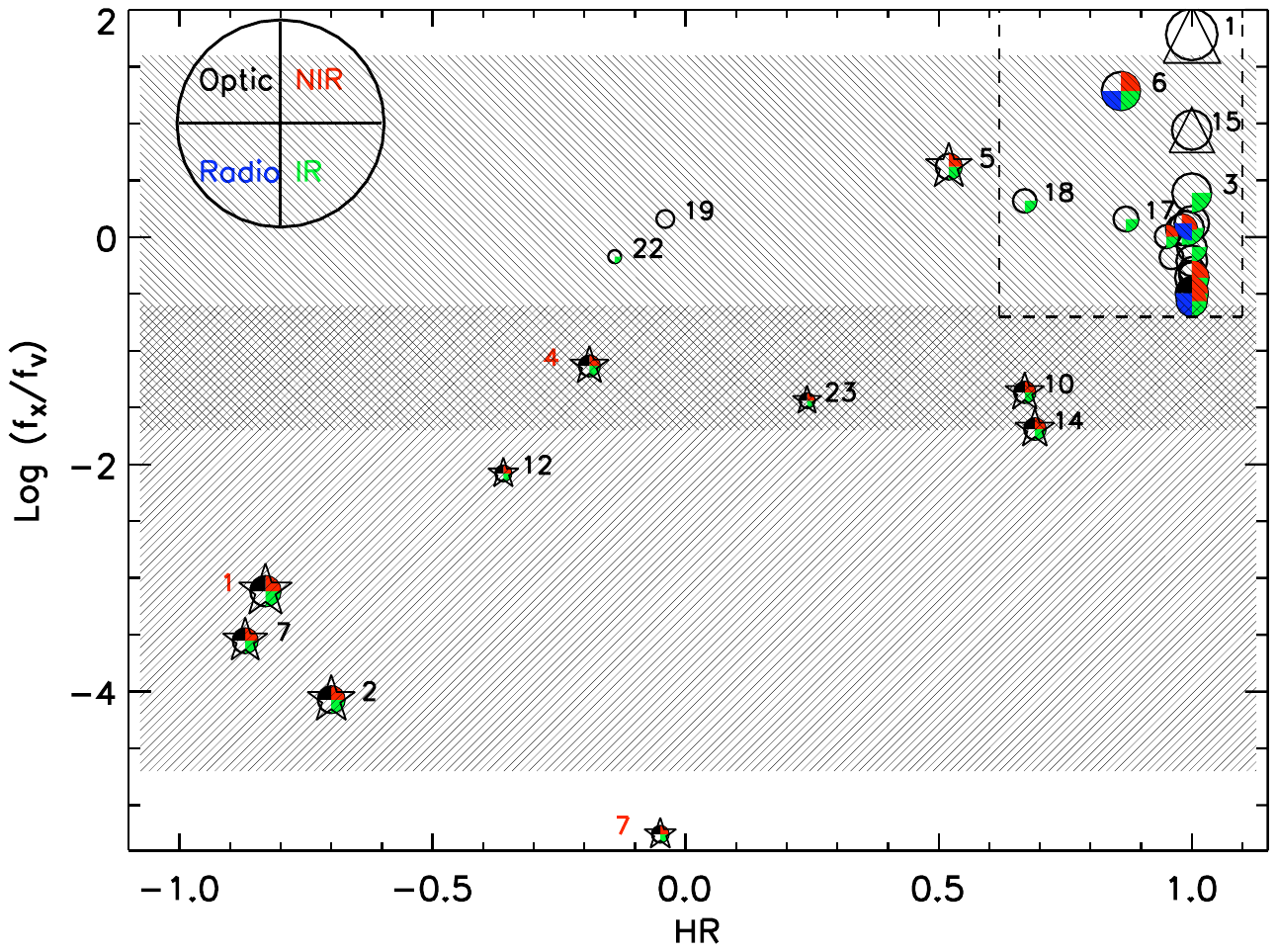}\vspace{-0.5cm}
\includegraphics[width=3.3in,angle=90]{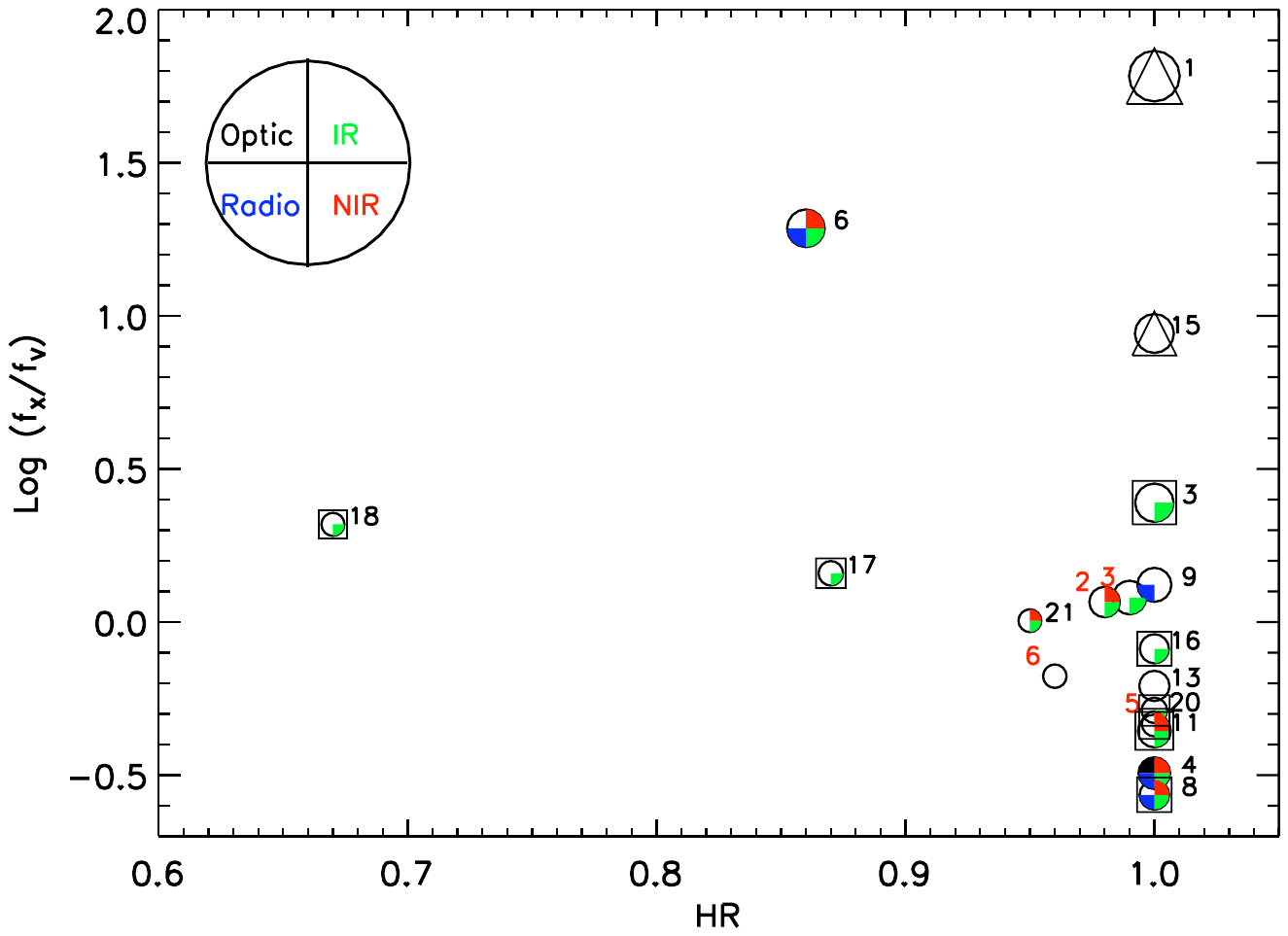}\vspace{-0.5cm}

\caption{ \footnotesize{Logarithm of the X-ray to optical flux ratio ($\log f_X/f_V$) versus hardness ratio (HR) for the X-ray sources detected in G23.5+0.1 and G25.5+0.0 fields  (denoted by red and black numbers, respectively, 
in accordance with Tables 2 and 3).  The bottom panel is the zoom-in of the upper right corner  
of  the top panel (shown by the dashed lines). The sizes of  the circles are proportional to  $\log f_X$.  The optical, 2MASS, GILMPSE, and radio detections are shown by 
 filling the circle quadrants with different colors, according to the legend shown in the upper left corner. An empty (white) quadrant means nondetection in the corresponding wavelength range. The circles with empty upper-left quadrants show lower limits on $\log f_X/f_V$.
Sources classified as foreground MS stars are additionally marked  by the ``star'' symbol, while those classified as an AGN or PMS-star are are marked by the ``square'' symbol. The young pulsar and the PWN candidate are marked by the ``triangle'' symbols. Classification of the sources shown by circles only (without additional symbols)  is uncertain (see Tables 6 and 7).
 The  shaded regions correspond to main sequence stars (B-M classes; bottom region) and extragalactic sources (galaxies and AGNs; top region) according to the criteria defined by Maccacaro et al.\ (1988). }}

\end{figure}

\begin{figure}[]

 \centering

\includegraphics[width=3.5in,angle=90]{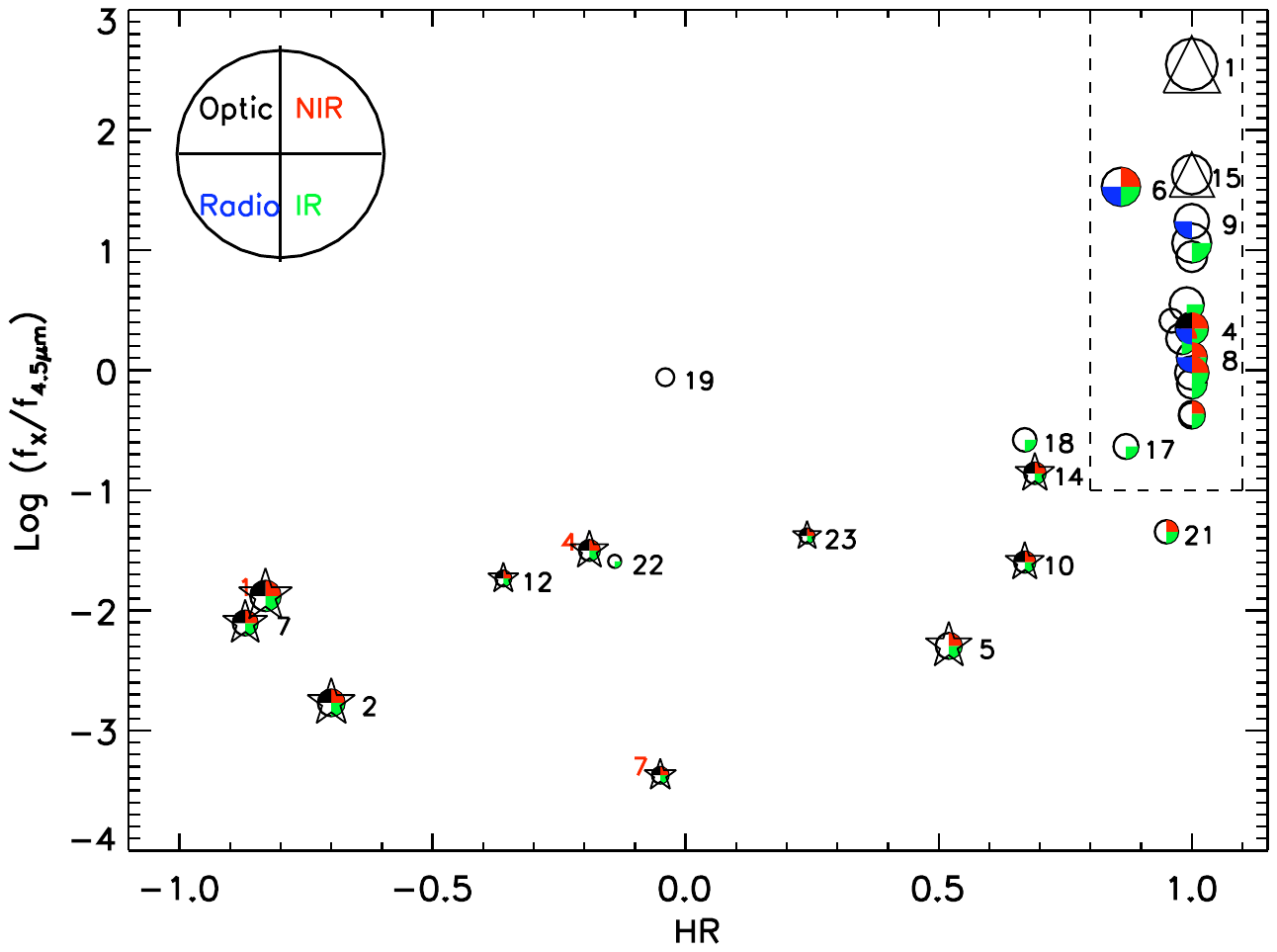}\vspace{-0.5cm}
\includegraphics[width=3.5in,angle=90]{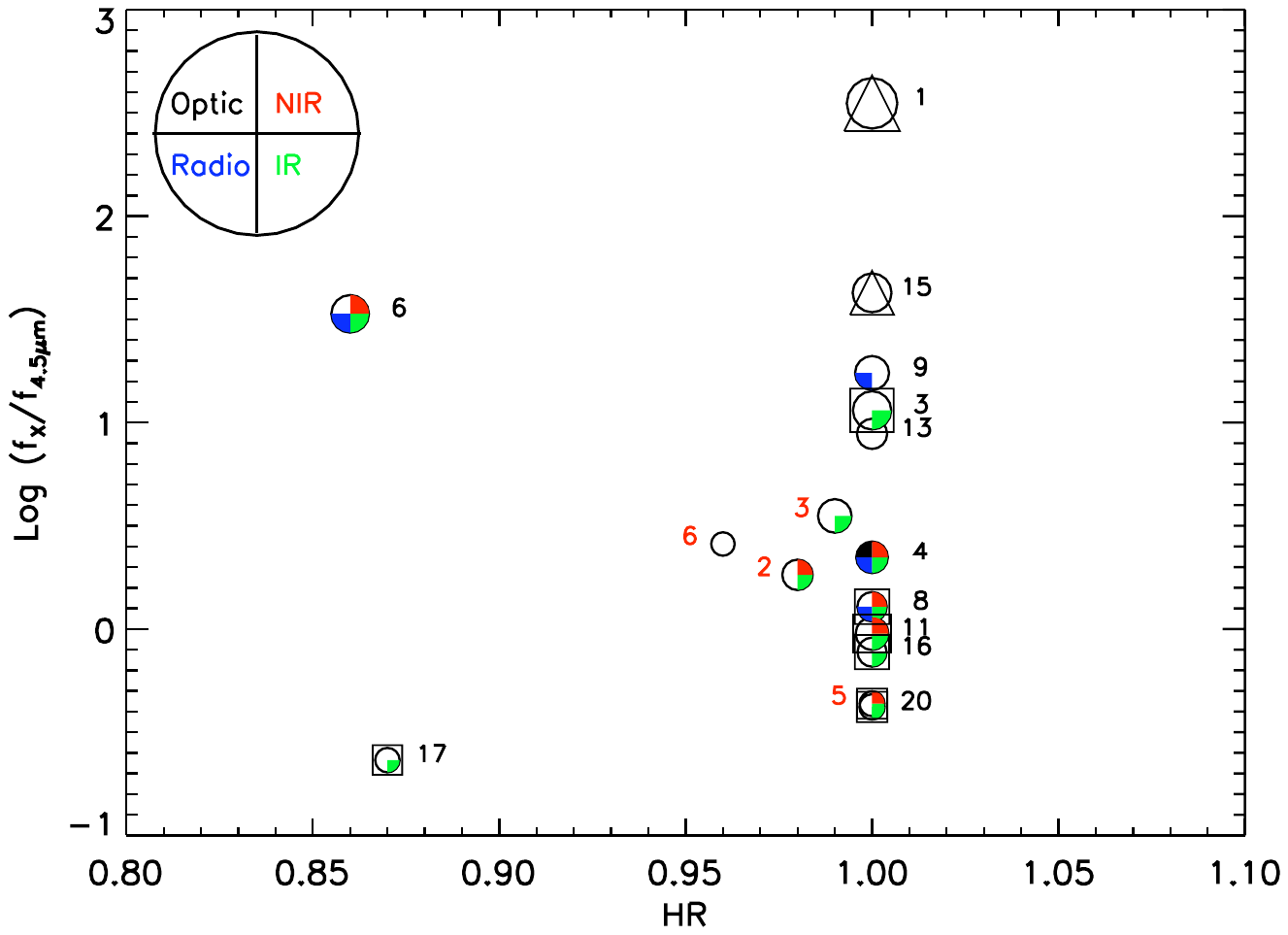}\vspace{-0.8cm}

\caption{ \footnotesize{ Logarithm of the X-ray to 4.5 $\mu$m flux ratio ($\log f_X/f_{4.5\mu{\rm m}}$) versus hardness ratio (HR) for X-ray sources detected in G23.5+0.1 and G25.5+0.0 fields (denoted by red and black numbers, respectively, 
in accordance with Tables 2 and 3).  The bottom panel is the zoom-in of the upper right corner  
of  the top panel (shown by the dashed lines). Sizes of  the circles are proportional to $\log f_X$.  The optical, 2MASS, GILMPSE, and radio detections are shown by 
 filling the circle quadrants  with different colors according, to the legend shown in the upper left corner. An empty (white) quadrant means nondetection in the corresponding wavelength range. The circles with empty upper-right quadrants show lower limits on $\log f_X/f_{4.5\mu{\rm m}} $ . Sources classified as foreground MS stars are additionally marked  by the ``star'' symbol, while those classified as an AGN or PMS-star are are marked by the ``square'' symbol. The young pulsar and the PWN candidate are marked by  the ``triangle'' symbols. Classification of the sources shown by circles only (without additional symbols)  is uncertain (see Tables 6 and 7). } }

\end{figure}

\end{document}